\documentclass[preprint,review,12pt]{elsarticle}

\usepackage{amssymb}

\usepackage[draft]{todonotes}   

\usepackage{url}
\usepackage{float}

\usepackage{graphicx}
\usepackage{dcolumn}
\usepackage{bm}
\usepackage[mathlines]{lineno}
\usepackage{color}
\usepackage[caption=false]{subfig}
\usepackage{multirow}
\usepackage{amsmath}
\usepackage{algorithmic}
\usepackage{paralist}

\DeclareMathOperator{\Tr}{Tr}
\DeclareGraphicsExtensions{.pdf}
\graphicspath{{figs/pdf/}}

\journal{Physica A}

\begin{document}

\begin{frontmatter}

\title{Assessment of Models for Pedestrian Dynamics with Functional Principal Component Analysis}
\author[label1]{Mohcine Chraibi\corref{cor1}}
\address[label1]{J\"ulich Supercomputing Centre,  Forschungszentrum J\"ulich GmbH, 52428 J\"ulich, Germany}
\cortext[cor1]{corresponding author}
\ead{m.chraibi@fz-juelich.de}
\author[label1wup]{Tim Ensslen}
\address[label1wup]{Department of Mathematics and Computer Science, Bergische Universit\"at Wuppertal, 42119 Wuppertal, Germany}
\ead{tim.ensslen@uni-wuppertal.de}
\author[label1wup]{Hanno Gottschalk}
\ead{hanno.gottschalk@uni-wuppertal.de}
\author[label1wup]{Mohamed Saadi}
\ead{saadi@uni-wuppertal.de}
\author[label1,label2wup]{Armin Seyfried}
\ead{a.seyfried@fz-juelich.de}

\address[label2wup]{Department of Civil Engineering, Bergische Universit\"at Wuppertal, Pauluskirche 7, 42285 Wuppertal, Germany}

\begin{abstract}
  Many agent based simulation approaches have been proposed for pedestrian flow. As such models are applied e.g.\ in evacuation studies, the quality and
  reliability of such models is of vital interest. Pedestrian trajectories are functional data and thus functional principal component analysis is a
  natural tool to asses the quality of pedestrian flow models beyond average properties. In this article we conduct functional PCA for the trajectories of
  pedestrians passing through a bottleneck. In this way it is possible to asses the quality of the models not only on basis of average values but also by considering its fluctuations.
  We benchmark two agent based models of pedestrian flow against the experimental data  using PCA average and stochastic features. Functional PCA proves
  to be an efficient tool to detect deviation between simulation and experiment and to asses quality of pedestrian models.
\end{abstract}


\end{frontmatter}

\section{\label{sec:introduction}Introduction}

Most of force-based models \textit{qualitatively} describe the movement of
crowds of pedestrians. Self-organization phenomena e.g., lane formations~\cite{Helbing1995,Helbing2004,Yu2005},
oscillations at bottlenecks~\cite{Helbing1995,Helbing2004},
clogging at exit doors~\cite{Helbing2004,Yu2005} etc.,  are reproduced.
From a physical point of view it is of interest how simple model reproduce qualitatively self-organization phenomena of driven multi-particle systems.
That contributes to a better understanding of the investigated systems and the essential interactions.
In addition numerical simulations basing of these models are used to address safety related issues,
concerning e.g.~design and conception of escape routes in buildings~\cite{Schneider2002,TraffGo2005} or optimal organization of mass events or public transport facilities (VISWalk~\cite{ptv}, Legion~\cite{legion}, $\ldots$). For such utilization a thorough \textit{quantitative} validation of the models is obligatory to ensure a reliable layout, dimensioning or evaluation of pedestrian facilities.
In most known cases this is fulfilled  by
reproducing the fundamental diagram~\cite{Seyfried2006,Nishinari2006,Chraibi2010a,Jian2010}  or measuring
the flow through bottlenecks~\cite{Hoogendoorn2006,Chraibi2010a,LiaoW2014b}.
An overview of quantitative validation of models by means of the fundamental diagram is given in~\cite{Seyfried2008}.
On one hand, the common point between these quantitative
methods is the fact that they are based on calculating  specific
traffic quantities, e.g.\ density, flow and velocity. On the other
hand, these measurements are performed based on locally averaged values over
time or space.
~\cite{Zhang2011a} and~\cite{Seyfried2010} provide examples how the measurement methods could influence the resulting empirical relations of such granular
and heterogeneous systems of finite size.
The differences between the measurement methods suggest that important information on the system may be lost during the measurement process.
Moreover state of the art models describe pedestrian dynamics on a more detailed level by simulating trajectories of every single pedestrian allowing in
principle a validation method assessing average pedestrian or traffic flow behavior, but also accounting for the amount and typical nature of fluctuation
around this average.

A first methodology based on exploiting information of individual trajectories was introduced in~\cite{Johansson2007} to calibrate the social force model.
While one pedestrian was moved according to the model the others were moved according to real trajectories. By means of an evolutionary algorithm the
deviations of the resulting trajectories from the experimental ones was used to calibrate the parameters of the model. But this approach doesn't allow an
assessment of the quality of a model. 

While an abundance of agent-based models in the field of pedestrian and traffic dynamics were developed in the last years~\cite{Chraibi2015,ped14}
the question of systematic comparison of experimental evidence and
model generated results has not caught the same attention.
This would however be important for the ranking of models into more or less adequate ones.
As argued above methodology of the evaluation should provide a comparison of model results and empirical data corresponding to the level of detail of the model.
It is  desirable that such a validation method should not only be able to asses average pedestrian or traffic flow behavior, but also account for the amount and typical nature of fluctuation around this average.

Among the difficulties in this validation process is the fact that in agent-based pedestrian or traffic flow data is functional, i.e.\ to each individual  we associate
data in the infinite dimensional space of trajectories $x(t)$.
The adequate statistical approach for the study of pedestrian or traffic flow data is thus the well established method of functional data
analysis~\cite{RS}.
In this method, the variation in the trajectories of different agents is interpreted as random fluctuations.
Thus, the measured or simulated trajectories are interpreted as realizations of some stochastic process $X(t)\in\mathbb{R}^2$, where $t$ stands for a time parameter and
$X(t)=X(t,\omega)$ tacitly  depends on some random parameter $\omega$ from a probability space $(\Omega,{\cal A},P)$. For more details the reader is  conferred
to~\cite{Bauer}.
Although there are infinitely many trajectories available for an agent to move from point $A$ to point $B$, it often turns out that a few typical modes of variation
around the average movement are responsible for the bulk of fluctuation of trajectories between different individuals.
As a classical method in the analysis of functional data, the functional principal component analysis (PCA) is the standard method to find and analyze these typical
variations.

The scope of this article is to use functional PCA analysis to study the performance of agent-based models of pedestrian motion with respect to experimental data.
In order to demonstrate the methodological approach, two models -- social force model (SFM)~\cite{Molnar1995} and generalized centrifugal force model
(GCFM)~\cite{Chraibi2010a} -- are used to simulate pedestrian movement through a bottleneck of the same dimensions.
In the following we apply functional PCA using the open source extension {\tt fda}  by Ramsey, Hooker and Graves~\cite{RHG}
to conduct the analysis.
We present the results and give a detailed comparison of average values for locations and velocities and their respective principal components.
For the latter we separately compare strength, distribution of total variation, and morphology of principal components.

We show that functional PCA in fact can be used to make statistically significant statements about model quality.
Functional PCA reveals significant deviation between both models and the experiment already on the level of average values.
While the morphology of principal components for locations is more or less adequately represented by both models, there are significant deviations in the strength of
fluctuations around the mean behavior with the GCFM model underestimating the experimentally observed fluctuations while the SFM mostly overestimates fluctuation
strength.
These empirical observations can be confirmed with statistical testing for significance using the PCA-bootstrap
methodology~\cite{DiaconisEfron1983,FisherCaffoSchwartzZipunnikov2014}.

In this article, for the fist time we combine functional PCA in the sense of~\cite{RS} with the bootstrapping of scores in
order to calculate the fluctuations of specific statistics that describe and distinguish characteristic features of
fluctuations of individual pedestrian behavior in a crowd.
Also on the PCA-side, benchmarking and testing with specific statistics evaluated in functional PCA is a new strategy, to the best of our knowledge.

The article is organized as follows.
In section~\ref{sec:experiment} we review the pedestrian flow experiment~\cite{Seyfried2010a} as the benchmark case for this study.
Section~\ref{sec:models} gives a brief account on the SFM and the GCFM model.
In Section~\ref{sec:pca}  reviews the functional PCA and its numerical implementation.
Section~\ref{sec:analysis} is the main part of this article.
After some introductory remarks on data formatting and smoothing (Subsection~\ref{sec:prep}),  we compare average data for
$x$ and $y$ position data (Subsection~\ref{sec:average}) and velocities directed in the main direction of motion, which is the $x$-direction.
We then compare fluctuations strength via PCA eigenvalues (Subsection~\ref{sec:pcastrength})
and morphology for the first PCA harmonics for $x$- and $y$- positions and $x$-velocities (Subsection~\ref{sec:pcaxy}).
Section~\ref{sec:StatInf} presents the PCA-bootstrap approach in the context of spline-based PCA
(Subsection~\ref{sec:BootMeth}) and applies this to total variation and Gini index (Subsection~\ref{sec:GiniTest}) as well as
the $L^2$-distance of the average trajectories and the Hilbert Schmid distance of the empirical correlation functions
(Subsection~\ref{sec:L2HSTest}).
In Section~\ref{sec:summary} we summarize or findings and give some conclusions on model quality in the specific case and
general applicability of functional PCA in the given context.

\section{\label{sec:experiment}Experiment}
\begin{figure}[htb]
  \centerline{
    \includegraphics[width=0.5\textwidth]{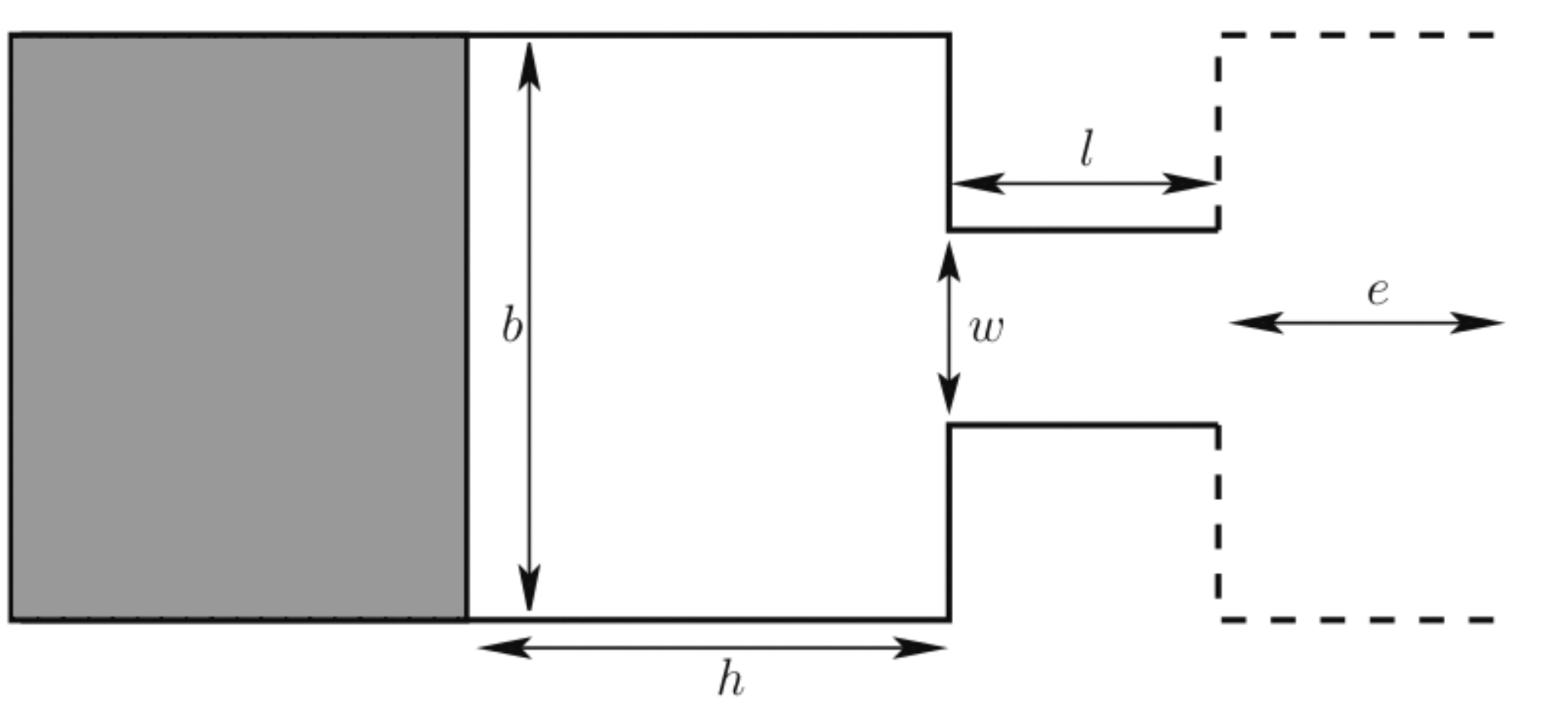}}
  \caption{The simulation set-up: pedestrians start from the shaded area and move
    through the bottleneck ($l=4$\, m, $h=4.5$\, m, $b=6$\, m and $w=0.9$\, m).
    An adjacent area of length $e=2.5$\, m is added to consider the backward
    effect of leaving pedestrians on those still in the bottleneck.}\label{fig:bot}
\end{figure}

In this work we use as a reference the experimental data extracted
from the experiment~\cite{Seyfried2010a}, that  was performed in 2006 in the wardroom of the ``Bergische
Kaserne D\"usseldorf''.
See Fig.~\ref{fig:bot}.

A waiting area was used to distribute the attendees before the start of each run of the experiment.
For simulation purposes we  enlarge the area of the set-up by
an extra room of length $e$. This is necessary to take into consideration the effects of pedestrians that leaved the bottleneck on
the pedestrians still in the system.

The flow through the bottleneck is measured as follows:
\begin{equation}
  J = \frac{N}{\Delta t},
\end{equation}
with $N$ the number of pedestrians and $\Delta t=t_{\rm last}-t_{\rm
  first}$ the time gap between the first and the last pedestrian
passing the bottleneck at the measurement line.

\section{\label{sec:models}Models}
Force-based models describe the movement of pedestrians as a superposition of
forces. Given the state variables of pedestrian $i$ at
time $t$ $(\overrightarrow{x_i}(t), \overrightarrow{v_i}(t))$ and considering Newton's second law of dynamics the state of each
pedestrian $i$ is defined by:
\begin{equation}
  m_i\overrightarrow{a_i}(t) =  \sum_{i\neq j}
  \overrightarrow{f_{ij}^{r}} + \sum_{W} \overrightarrow{f_{iw}^{r}}
  + \overrightarrow{f_{i}^{d}},
  \label{eq:maineq}
\end{equation}
and
\begin{equation}
  \overrightarrow{v_i}(t) = \frac{d\overrightarrow{x_i}(t)}{dt},
\end{equation}
where $\overrightarrow{f_{ij}^{r}}$ denotes a repulsive force
acting from the $j^{th}$-pedestrian on the $i^{th}$-pedestrian,
$\overrightarrow{f_{iw}^{r}}$ is a repulsive force emerging from
borders, walls etc.\  and $\overrightarrow{f_{i}^{d}}$ is a driven force. $m_i$
is the mass of pedestrian $i$. 

The superposition of the forces
reflect the fact that pedestrians move towards a certain point in
space (e.g.\ an exit) and meanwhile try to avoid collisions with each
other or with walls and objects.

The driving force $\overrightarrow{f_{ij}^{r}}$ models, at low densities, an exponential acceleration towards a desired speed $v_0$:
The following expression~\cite{Pipes1953}  is used:
\begin{equation}
  \overrightarrow{f_i^d} = m_i \frac{ v_0\overrightarrow{e_i^0} -
    \overrightarrow{v_i} } {\tau},
  \label{eq:fdrv}
\end{equation}
with a relaxation time $\tau$ typically equal to 0.5 s, and a desired
direction $\overrightarrow{e_{i}^0}$ of pedestrian $i$.

The repulsive force between pedestrians  $\overrightarrow{f_{ij}^r}$ is defined
differently from one model to another
\cite{Helbing1995,Seyfried2006,Yu2005,Chraibi2010a,Shiwakoti2011,Karamouzas2014}.

\begin{figure}[htb]
  \centerline{
    \includegraphics[width=0.4\textwidth]{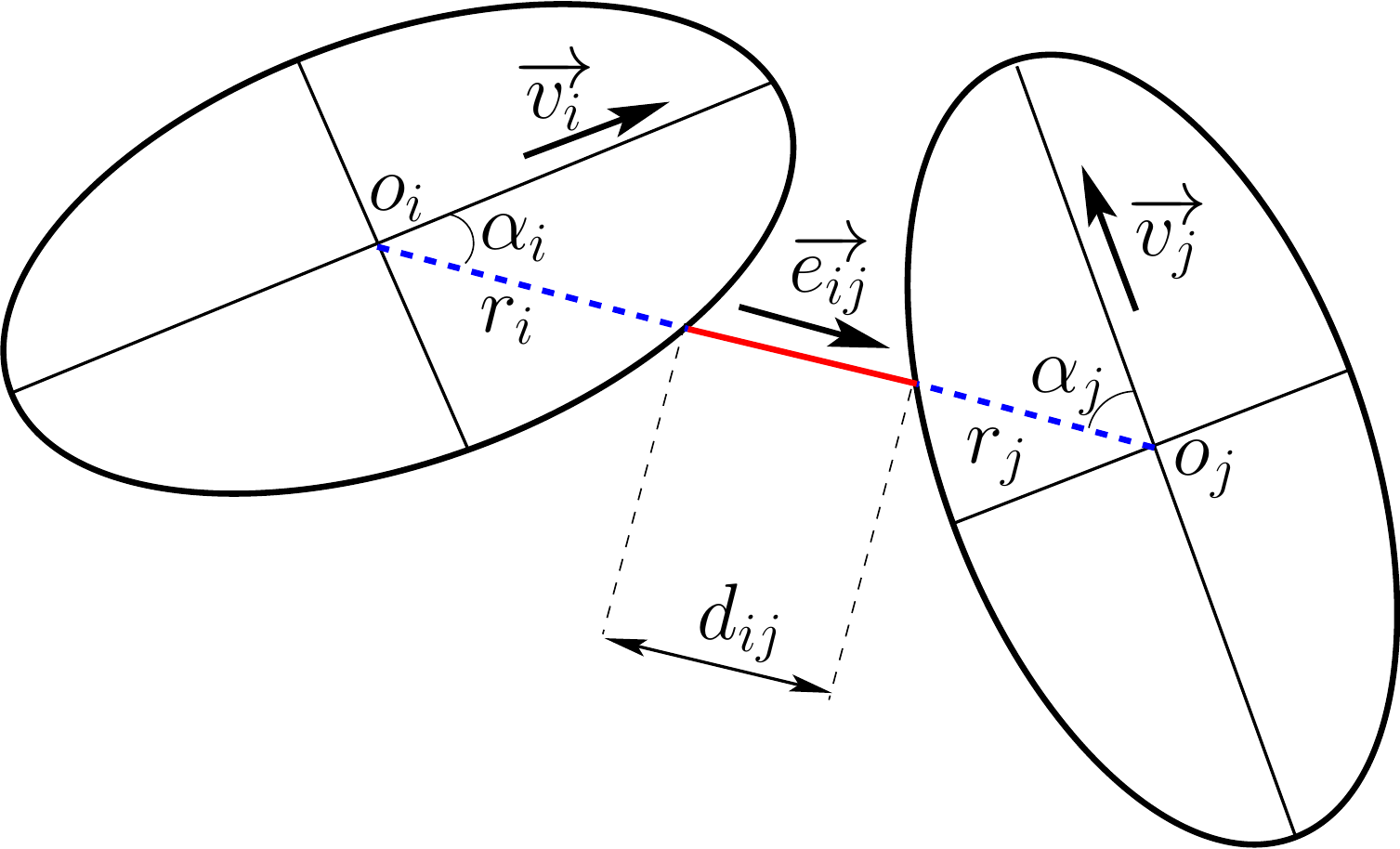}}
  \caption{The effective distance $d_{ij}$ of two pedestrians
    represented by two ellipses.}
  \label{fig:dist}
\end{figure}

In this work we study a variation of the SFM  and the GCFM.
Both models are microscopic and continuous in space. In the GCFM  the agents have an elliptical form with velocity-dependent semi-axes, whereas the shape of agents in
the SFM is circular. In the general case, the distance  $\parallel\overrightarrow{d_{ij}}\parallel$
is defined as the distance between  the borders of the ellipses
$i$ and $j$ along a line connecting their centers. See Fig.~\ref{fig:dist}. For the SFM the semi-axis orthogonal to the movement direction is equal to the other
semi-axis in the direction of movement.
For simplicity we write $d_{ij}$ to denote the norm of the vector  $\overrightarrow{d_{ij}}$.

\subsection{The social force model (SFM)}
\label{sec:sfm}
The SFM as originally published by Moln\'ar~\cite{Molnar1995} describes the movement of circular agents as superposition of different
factors e.g.\ influence of neighboring pedestrians, walls, attractions and groups. In this work we reduce the complexity of the model to a minimum, considering only the
influence of pedestrians and walls and assuming only circular potentials.

The repulsive force in the SFM  between agent $i$ and $j$ is defined as 
\begin{equation}
  \overrightarrow{f_{ij}^{r}} = -m_i k_{ij}
  A\exp(\frac{d_{ij} }{B}) \overrightarrow{e_{ij}}. 
  \label{eq:sfmfrep}
\end{equation}
with
\begin{equation}
  \overrightarrow{d_{ij}} = \parallel \overrightarrow{x_j} -\overrightarrow{x_i}\parallel - r_i - r_j,
  \label{eq:R}
\end{equation}
and
\begin{equation}
  \overrightarrow{e_{ij}} = \frac{\overrightarrow{x_j} -\overrightarrow{x_i} }{ \parallel \overrightarrow{x_j} -\overrightarrow{x_i}\parallel },
  \label{eq:relv}
\end{equation}
with the parameters $A$ and $B$ the strength and the range of the force are adjusted.
The limited vision of pedestrians ($180^\circ$) is modeled by the coefficient
$k_{ij}$:
\begin{equation}
  k_{ij} =\Theta\Big(\overrightarrow{v_i}.\overrightarrow{e_{ij}}\Big).
  \label{eq:K}
\end{equation}
$\Theta(\cdot)$ is the Heaviside function.
The repulsive force between pedestrians and static objects is defined
similarly to (\ref{eq:sfmfrep}).

\subsection{The generalized centrifugal force model (GCFM)}
The repulsive force in the GCFM is inversely proportional to the distance of two ellipses representing moving pedestrians $i$ and $j$ and depends on their relative velocity:

\begin{equation}
  \overrightarrow{f_{ij}^r}=-m_i\frac{\Big(\alpha v_0  +
    v_{ij}\Big)^2}{d_{ij}},
  \label{eq:gcfmfrep}
\end{equation}
where $ v_{ij} = \Theta\Big((\overrightarrow{v_i}-\overrightarrow{v_j})\cdot\overrightarrow{e_{ij}}\Big),\; $ is the relative velocity.
The use of the Heaviside function $\Theta(\cdot)$ ensures, that faster pedestrians are not effected by slower pedestrians.
By means of the parameter $\alpha$ the strength of the force can be adjusted.
As mentioned earlier in the GCFM the space requirement in the direction of movement is modeled by the semi-axis
\begin{equation}
  a= a_\text{min} + \tau_a v_i,
  \label{eq:a}
\end{equation}
with two parameters $a_\text{min}$\, and $\tau_{a}$, whereas  the lateral swaying of pedestrians is modeled by the semi-axis
\begin{equation}
  b= b_\text{max} - (b_\text{max}-b_\text{min})\frac{v_i}{v_0}.
  \label{eq:b}
\end{equation}
\subsection{Model parameters}
\begin{figure}[htb]
  \centerline{
    \includegraphics[width=0.6\textwidth]{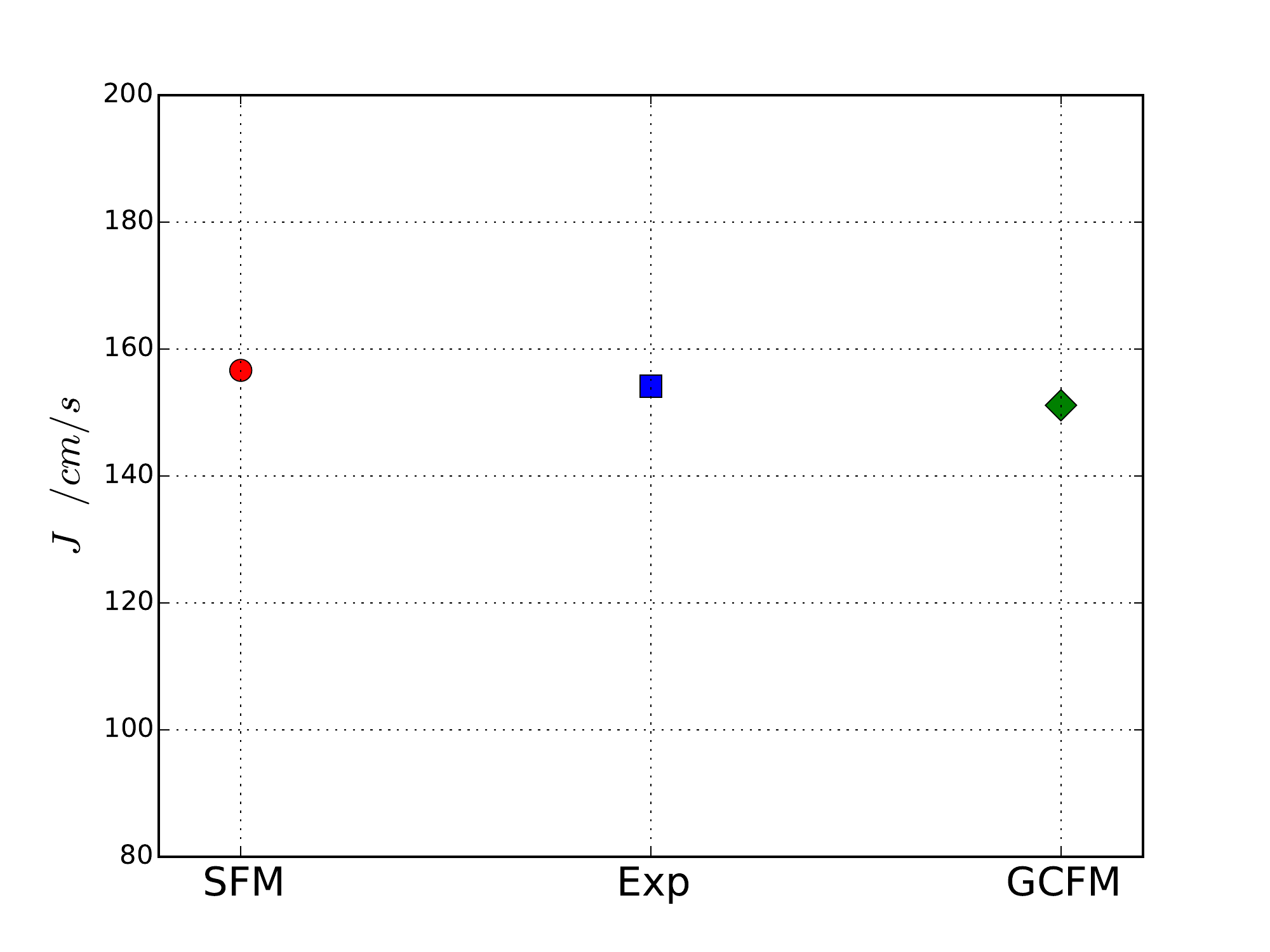}
  }
  \caption{The flow through the bottleneck measured at the middle of the corridor after the entrance to the bottleneck. The empirical value is a reference value for the calibration of the GCFM and the SFM.}
  \label{fig:flow}
\end{figure}

As mentioned earlier the original SFM includes several forces e.g.\ physical contact forces and attractive forces.
For our purpose we use a  simplified version of the SFM as presented in Sec.~\ref{sec:sfm}.
We choose $A=5$\, N for pedestrian-pedestrian interactions (\ref{eq:sfmfrep}) and $A=7$\, N for pedestrian-wall interactions.
The range of the function defined by the parameter $B$ in  (\ref{eq:sfmfrep}) was chosen to be $0.08$\, m for pedestrian-pedestrian interactions and  $0.05$\, m for
pedestrian-wall interactions.
The parameter $\alpha$ in (\ref{eq:gcfmfrep}) is set to 0.2 for pedestrian-pedestrian interactions and 0.33 for pedestrian-wall interactions.
The desired speed $v_0$ is set to $\mu=1.1$\, m/s.
For simplicity we set for both models $m_i=1$\, Kg. Table~\ref{tab:param} gives a resume of the  parameters used.
\begin{table}[H]
  \begin{center}
    \begin{tabular}{ lcrr }
      \hline
      Parameter & Equation & Value \\
      \hline
      $A_{\rm ped}$ & (\ref{eq:sfmfrep}) & 5 N\\
      $A_{\rm wall}$ & Similar to (\ref{eq:sfmfrep}) & 7 N\\
      $B_{\rm ped}$ & (\ref{eq:sfmfrep}) & 0.08 m\\
      $B_{\rm wall}$ & Similar to (\ref{eq:sfmfrep}) & 0.05 m\\
      $\alpha_{\rm ped}$  & (\ref{eq:gcfmfrep})  & 0.2\\
      $\alpha_{\rm wall}$  & Similar to (\ref{eq:gcfmfrep})  &0.33\\
      $\tau$ & (\ref{eq:fdrv}) &0.5 s\\
      $v_0$ &  (\ref{eq:fdrv})  &1.1 m/s\\
      $m$ &  (\ref{eq:fdrv})  & 1 Kg\\
      $a_\tau$ & (\ref{eq:a}) & 0.12 s \\
      $a_{\rm min}$ & (\ref{eq:a})  & 0.15 m \\
      $b_{\rm min}$ & (\ref{eq:b}) & 0.15 m\\
      $b_{\rm max}$ &  (\ref{eq:b}) & 0.2 m\\
      \hline
    \end{tabular}
  \end{center}
  \caption{Parameter values in simulations with both GCFM and SFM.}
  \label{tab:param}
\end{table}
The values chosen in Tab.~\ref{tab:param}  differ from the values published in other works~\cite{Lakoba2005, Chraibi2010a}.
Our choice of  the above mentioned values is  supported by qualitative reasons, ensuring minimal overlapping among pedestrians, as well by quantitative consideration of
the flow through the bottleneck.
See Fig.~\ref{fig:flow}.
For safety relevant simulations a careful calibration of the used models is needed. Having calibrated two different models based on usual qualitative and quantitative
criteria,  we strive to apply a new technique to assert the goodness of the investigated models and verify whether the aforementioned validation is sufficient to ensure
a trustworthy and safe use of the produced simulations.

\section{\label{sec:pca}Functional PCA:\ Foundations}

\subsection{What is functional PCA?}

In this section we give some details of functional PCA following~\cite{RS}.
The principal component analysis uses the principal axis transformation for multivariate, correlated numerical data using the (empirical) covariance information between
the single random variables.
Eigenvalues then sort the importance of the single eigenvectors (also called harmonics or modes) according to the variance. 

This concept needs to be adapted to the case where the observed data from $n$ individuals are functions -- as it is the case with the trajectories of pedestrians.
The variability of the data can still be described in with the eigenvalues and eigenvectors of the covariance function seen as an operator on the function space of
square integrable functions.
Given the stochastic process of random trajectories $X(t)$, with $t\in(0,L)$, the covariance function is defined as
\begin{equation}
  \label{eqa:Cov}
  C(s,t)=\mathbb{E}\left[(X(s)-\mathbb{E}[X(s)])(X(t)-\mathbb{E}[X(t)])\right],
\end{equation}
with $\mathbb{E}$ the expected value with respect to the underlying probability space. This covariance function needs to be estimated out of the data $x_j(t)$
\begin{equation}
  \label{eqa:CovEmp}
  \hat C(s,t)=\frac{1}{n-1} \sum_{j=1}^ n (x_j(s)-\bar x(s))(x_j(t)-\bar x(t)),
\end{equation}
with $x_j(t)$ the $j$-th observation is one realization  of the random process $X(t)$ and $\bar x(t)=\frac{1}{n}\sum_{j=1}^ n x_j(t)$.

In the following we assume that average values have already been removed from the stochastic signal, i.e.\ we consider transformed random quantities $X(t)\to X(t)-\mathbb{E}[X(t)]$ with estimated observations $x_i(t)-\bar x(t)$.
The eigenvalues of $C(s,t)$   can then be calculated after solving the follow eigenvalue equation
\begin{equation}
  \label{eqa:eigen}
  \int_0^LC(s,t)\xi(t)\, dt=\lambda \, \xi(s).
\end{equation}
This results in a set of eigenvalues $\lambda_1\geq\lambda_2\geq \cdots\geq 0$ and corresponding eigenfunctions $\xi_i(s)$. These eigenfunctions are orthonormal $\int_0^L\xi_i(t)\xi_j(t) \,dt=\delta_{i,j}$, where $\delta_{i,j}=1$ for $i=j$ and zero otherwise.
The eigenfunctions and eigenvalues can now approximately be determined from the observations $x_j(t)$ by replacing $C(s,t)$ by its empirical counterpart $\hat C(s,t)$.

\subsection{Numerical approximation}
The problem (\ref{eqa:eigen}) is an infinite dimensional eigenvalue problem and its empirical counter part is potentially very high dimensional (of dimension $n$).
A frequently used method to make this problem numerically tractable is to project the covariance matrix on the space spanned by some finite basis,  - e.g.\ a
sufficiently fine B-spline or Fourier basis. Then one solves for the eigenvalues and functions in the given finite dimensional space of basis functions.
Therefore we approximate the observed functions $x_i(t)$ with a suitable linear combination of basis functions
\begin{equation}
  \label{eqa:basis}
  x_i(t)=\sum_{k=1}^K c_{i,k}\Phi_k(t)\Leftrightarrow x_i(t)=\bf{C}\Phi(t),
\end{equation}
with basis function vector $\Phi(t)=(\Phi_1(t),\ldots,\Phi_K(t))^T$ and coefficient matrix ${\bf C}=(c_{j,k})_{j=1,\ldots,n\atop k=1,\ldots,K}$ obtained e.g.\ by
orthogonal projection of $x_i(t)$ to the space spanned by the basis functions and a subsequent basis decomposition, in which case ${\bf C}={\bf W}^{-1}{\bf v}_j$ with
$W_{i,j}=\left\langle\Phi_i,\Phi_j\right\rangle=\int_0^L\Phi_i(t)\Phi_j(t)\, ds$ and $c_{j,k}=\langle x_j,\Phi_k\rangle=\int_0^Lx_j(t)\Phi_k(t)\, dt$.
Here some numerical quadrature may be employed for the integrals involved in the definition of $v_{j,k}$, whereas in most cases analytic formulae are available for $W_{i,j}$.
This projection method implies that the covariance function can be approximated by
\begin{equation}
  \hat C(s,t)= \frac{1}{n-1}\Phi(s)^T{\bf \bar C}^T{\bf \bar C}\Phi(t).
\end{equation}
Here we use $\bar C$ for the matrix of coefficients of $x_j(t)-\bar x(t)$ with respect to the basis $\Phi(t)$, namely $\bar c_{j,k}=c_{j,k}-\frac{1}{n}\sum_{j=1}^nc_{j,k}$.
Now we expand the eigenfunction  with the same basis functions to a good approximation:
\begin{equation}
  \xi(t)=\sum_{k=1}^Kb_k\Phi_k(t)=\Phi(t)^T{\bf b}.
\end{equation}
The approximate eigen value equation can be written as
\begin{equation}
  \label{eqa:eigen2}
  \int_0^L\hat C(s,t)\xi(t)\, dt=\frac{1}{n}\Phi(s)^T{\bf \bar C}^T {\bf \bar C}{\bf W}{\bf b}=\lambda\Phi(s)^T{\bf b}=\lambda \xi(s). 
\end{equation}
leading to the eigenvalue equation ${\bf \bar C}^T {\bf \bar C}{\bf W}{\bf b}=\lambda{\bf b} \Leftrightarrow {\bf W}^{1/2}{\bf \bar C}^T {\bf \bar C}{\bf W}^{1/2}{\bf u}=\lambda{\bf u},~{\bf b}={\bf W}^{-1/2}{\bf u}$,
which can be solved numerically. The result is a number of eigenvalues $\lambda_1\geq \lambda_2 \geq \cdots\geq 0$ and coefficient vectors ${\bf b}_i$ for approximate
principal components $\xi_i(t)=\Phi(t)^T{\bf b}_i$, for $i=1,2,\ldots, n-1$, which is the maximal rank of ${\bf \bar C}$.

\subsection{Statistics of the eigenvalues}
In this subsection, we discuss how to reduce the information from the set of eigenvalues to a few significant characteristics.
In particular we will focus on code figures that measure the strength of fluctuations and their concentration to a few, active modes.

The eigenvalue $\lambda_i$ represents the strength of fluctuations in the respective mode of characteristic shape $\xi_i(t)$. The relative strength $\rho_i$ of the variation in the mode $\xi_i(t)$ and the cumulative relative strength $L_j$ up to the $j$-th mode $\xi_j(t)$ is given by
\begin{equation}
  \rho_i=\frac{\lambda_i}{\sum_ {j=1}^n \lambda_j} \mbox{ and } L_j=\sum_{i=1}^j\rho_i.
\end{equation}

Two quantities that can be derived form the eigenvalues of the PCA are of special interest: First, the total variation strength is simply the sum of all eigenvalues
$\Lambda=\sum_{j=1}^n\lambda_j$ whereas the Gini index is a measure of concentration that is build from the Lorenz curve quantities  $L_j$ via
$G=\frac{2}{n-1}\sum_{j=1}^n(L_j-j/n)$.
Geometrically, the Gini index measures the area between the diagonal and the Lorentz curve, cf.~e.g. Figure \ref{fig:strengthX} in the right panel.
It is normalized such that it takes the value one if only one mode is active and takes the value zero when all modes are equally activated
$\lambda_1=\lambda_2=\cdots=\lambda_n$.
Note that the order of $\lambda_j$ is descending in contrast to the usual definition of the Gini index, where the order is ascending.
As an alternative, one could also consider the entropy of the distribution of the total activity to the single modes.
The result of the observation however remain largely unchanged.

\subsection{\label{sec:Deviation}Deviation measures}
In this section we derive some quantities that can be used to measure the distance between one set of functional data and another such data set.
In particular we will utilize these distances for benchmarking models with respect to their distance to the experiment.
Two distance measures will be employed in the following: First, the mean quadratic deviation between the average trajectories of the model on the one hand and the experimental data on the other.
Secondly, we consider the Hilbert-Schmidt norm between the respective empirical covariance functions as a measure of the distance of the fluctuation behavior of the experiment and the simulation.
In the following we work with the data after projection to a finite spline basis $\Phi_k(t)$.

We start with the mean quadratic difference in the average behavior.
The mean of the observed function $x_i(t)$ is:
\begin{equation}
  \overline{x}(t) = \frac{1}{n}\sum_{j=1}^{n} x_j(t) = \sum_{k=1}^{K}\biggl(\underbrace{\frac{1}{n}\sum_{j=1}^{n}c_{jk}}_{=C_k}\biggr)\Phi_k(t) = \sum_{k=1}^{K}C_k\Phi_k(t).
\end{equation}
The mean quadratic distance, the squared $L^2$ norm,  of the difference between $\overline{x}(t)=\sum_{k=1}^{K}C_k\Phi_k(t)$ and $\overline{y}(t)=\sum_{k=1}^{K}C_{k}^{'}\Phi_k(t)$ is:
\begin{equation}
  \|\overline{x}(t) - \overline{y}(t) \|_{2}^{2} =\left\langle\sum_{k=1}^{K}(C_k-C_{k}^{'})\Phi_k(t),\sum_{k=1}^{K}(C_k-C_{k}^{'})\Phi_k(t)\right\rangle
  = ({\bf C}-{\bf C}')^T{\bf W}({\bf C}-{\bf C}').
\end{equation}
We now derive formulae for measuring the distance between experiment and simulation in the covariance structure.  
Let  $\hat D(s,t) = \hat C(s,t) - \hat C'(s,t) = \sum_{k=1}^{K}\sum_{l=1}^{K}(\underbrace{\bar{c}_{k,l} - \bar{c}'_{k,l}}_{=d_{k,l}}  )\Phi_k(t)\Phi_l(s)$ be the difference
of covariance functions. The Hilbert-Schmidt norm of $\hat D(s,t)$ is:
\begin{equation}
  \|\hat D\|_{HS}^2 =\int_0^L\int_0^L\left(\hat D(s,t)\right)^2dsdt=\Tr\bigl((\bf{D}{\bf W})^T\bf{D}{\bf W} \bigr).
\end{equation}
Here $\Tr(A)$ stands for the trace of the matrix $A$ and $\bf{D}$ is the matrix with entries $d_{j,k}$.

\section{\label{sec:analysis}PCA Results}

In this section we show that functional PCA is a useful tool for detailed validation of models for pedestrian dynamics.
Ideally, the variability in the data can be described with the help of the PCA with a few principal components.
These main components can be interpreted as the schemes for the deviation of individual trajectories from the mean flow.
This allows a comparison of simulated and experimental data beyond averaged flow features.
Therefore, we apply the PCA to the experimental and simulated data from two models SFM and GCFM and compare the results.
Here we apply the PCA for $x$ and $y$ coordinates over time separately, as this approach is somewhat more accessible to the interpretation.
For the alternative approach of jointly analysing $x$ and $y$ trajectories and a discussion of the pros and cons of both approaches, see~\cite{RS,RHG}.

The analysis of the data is based on the {\tt R} package {\tt fda} developed by J.O. Ramsay et al.~\cite{RHG} with some minor extensions by the authors.

\subsection{\label{sec:prep}Preparation of the Data}
\begin{figure}[htb]
  \centerline{
    \includegraphics[width=0.32\textwidth]{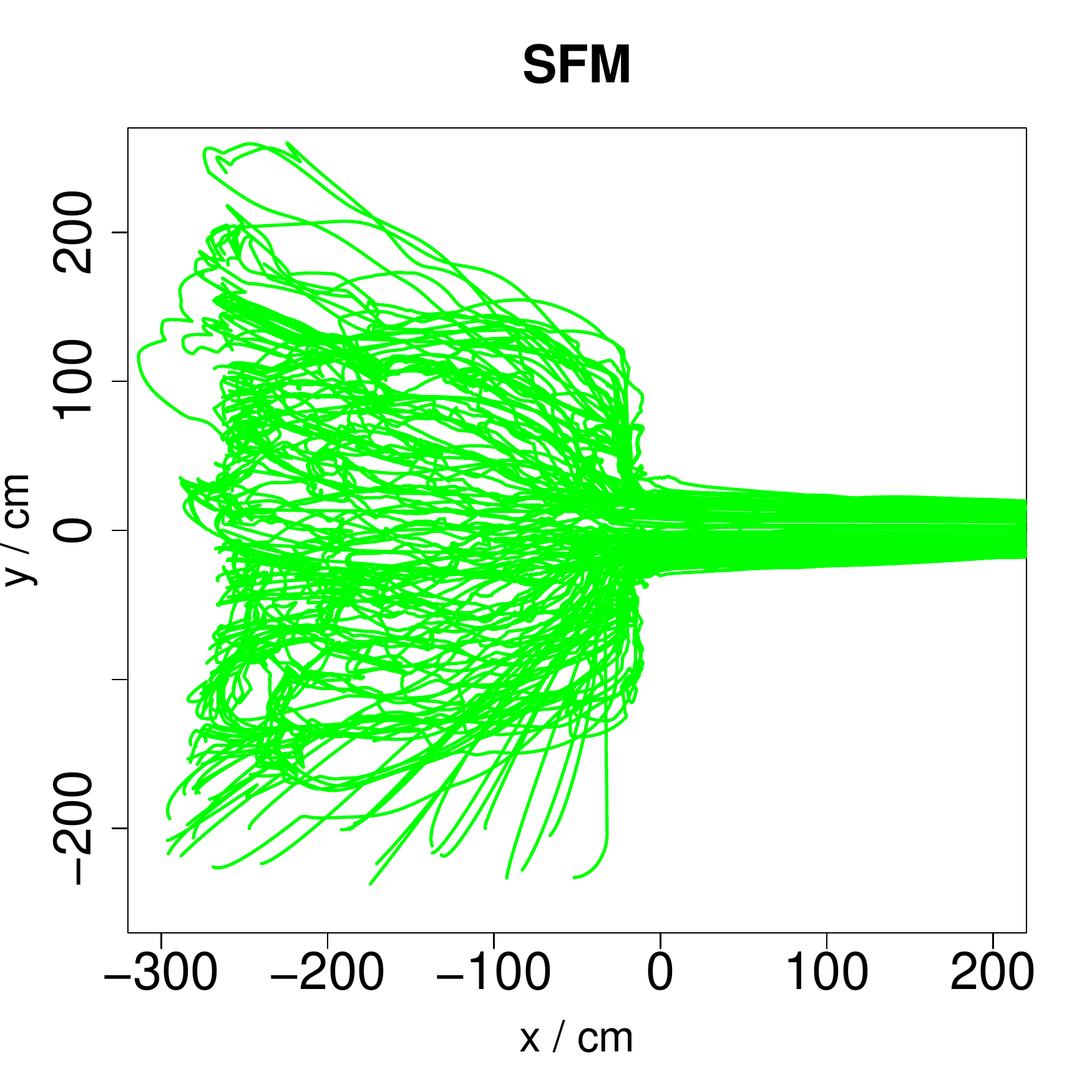}
    \includegraphics[width=0.32\textwidth]{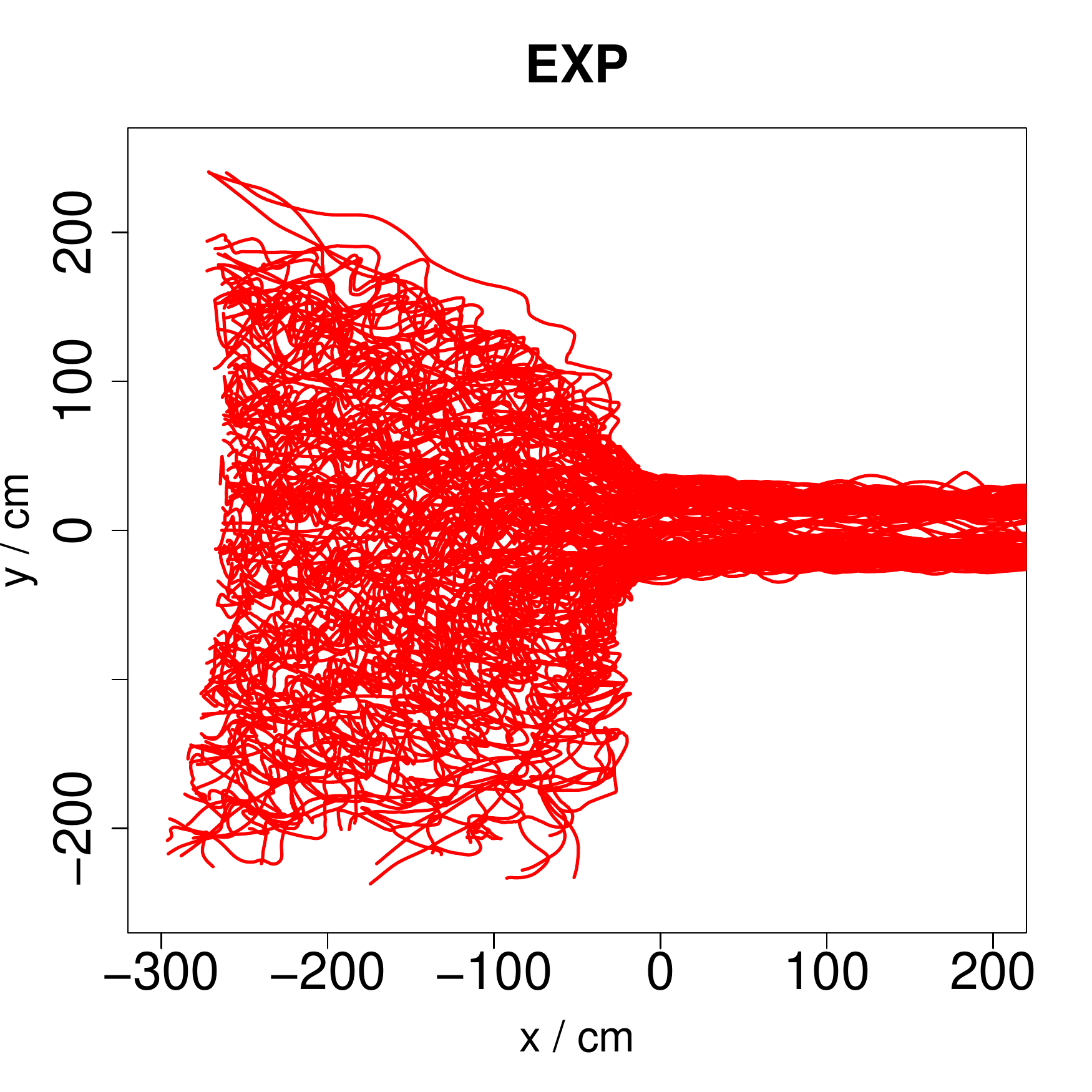}
    \includegraphics[width=0.32\textwidth]{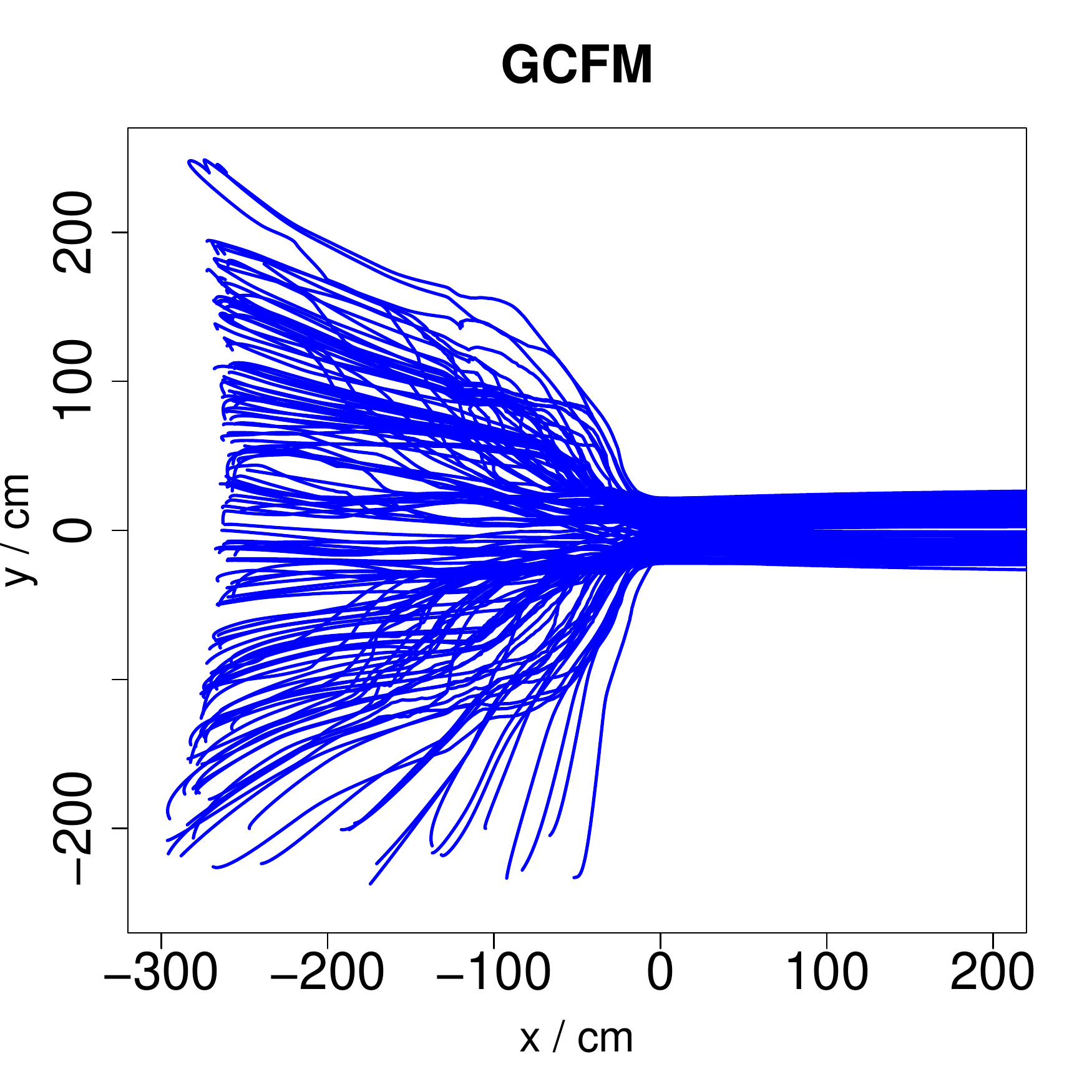}
  }
  \caption{XY plots of pedestrian trajectories generated by the SFM model (left), experiment (middle) and GCFM model (right).}
  \label{fig:XYplots}
\end{figure}

The pedestrian trajectory data in the experiment is recorded electronically with video tracking at the rate of 25 frames per second.
A total number of 149 trajectories has been recorded.
Likewise, the SFM and GCFM models have been simulated with 25 time steps per second using Euler integration with a time step  $\Delta t = 0.01$\, s.
For both models, a total of 149 trajectories have been simulated. See the trajectories in Figure~\ref{fig:XYplots}.

For the analysis, the pedestrian motion has to be stationary, i.e.\ all agents move under the same conditions.

\begin{figure}[htb]
  \centering
  \includegraphics[width=0.32\textwidth]{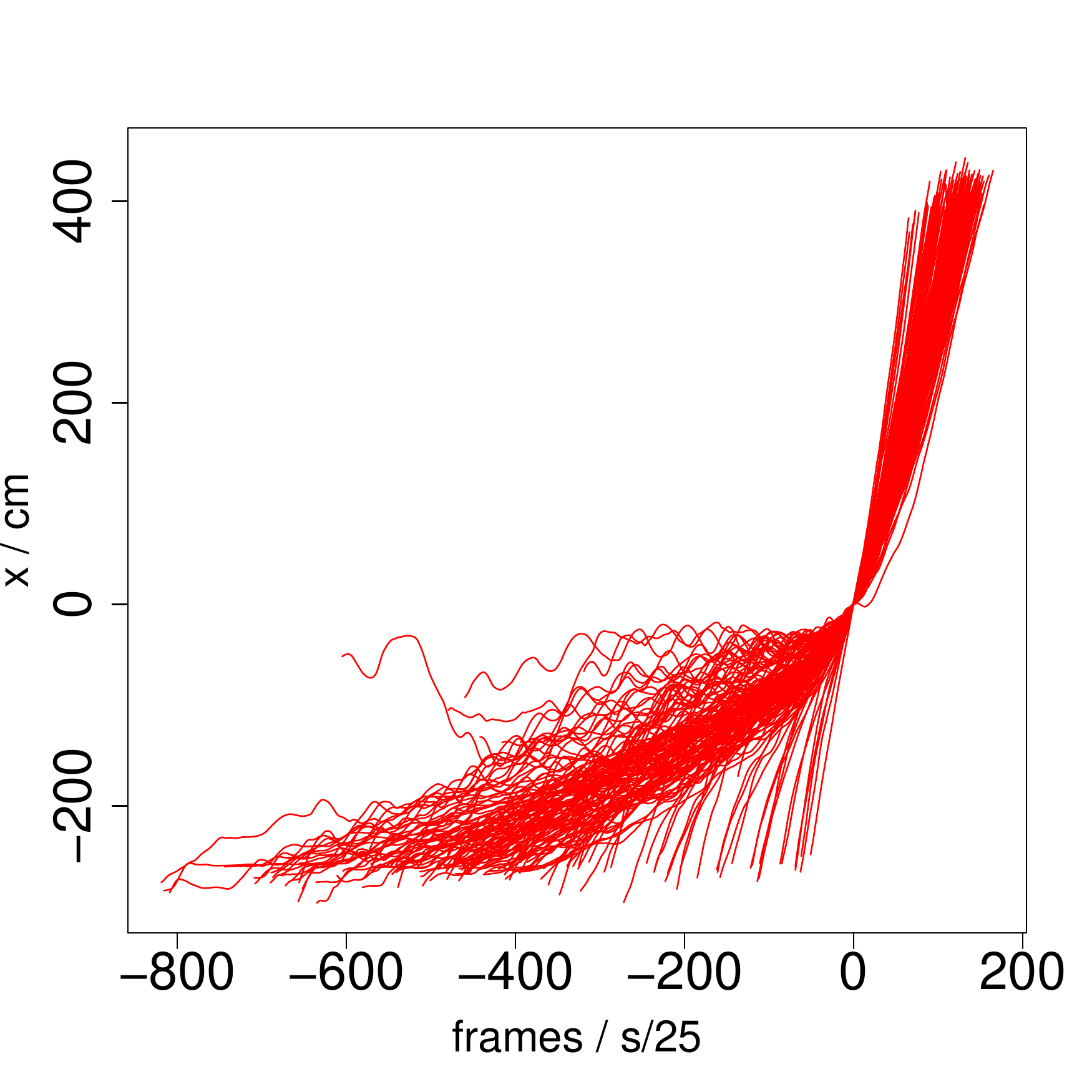}
  \includegraphics[width=0.32\textwidth]{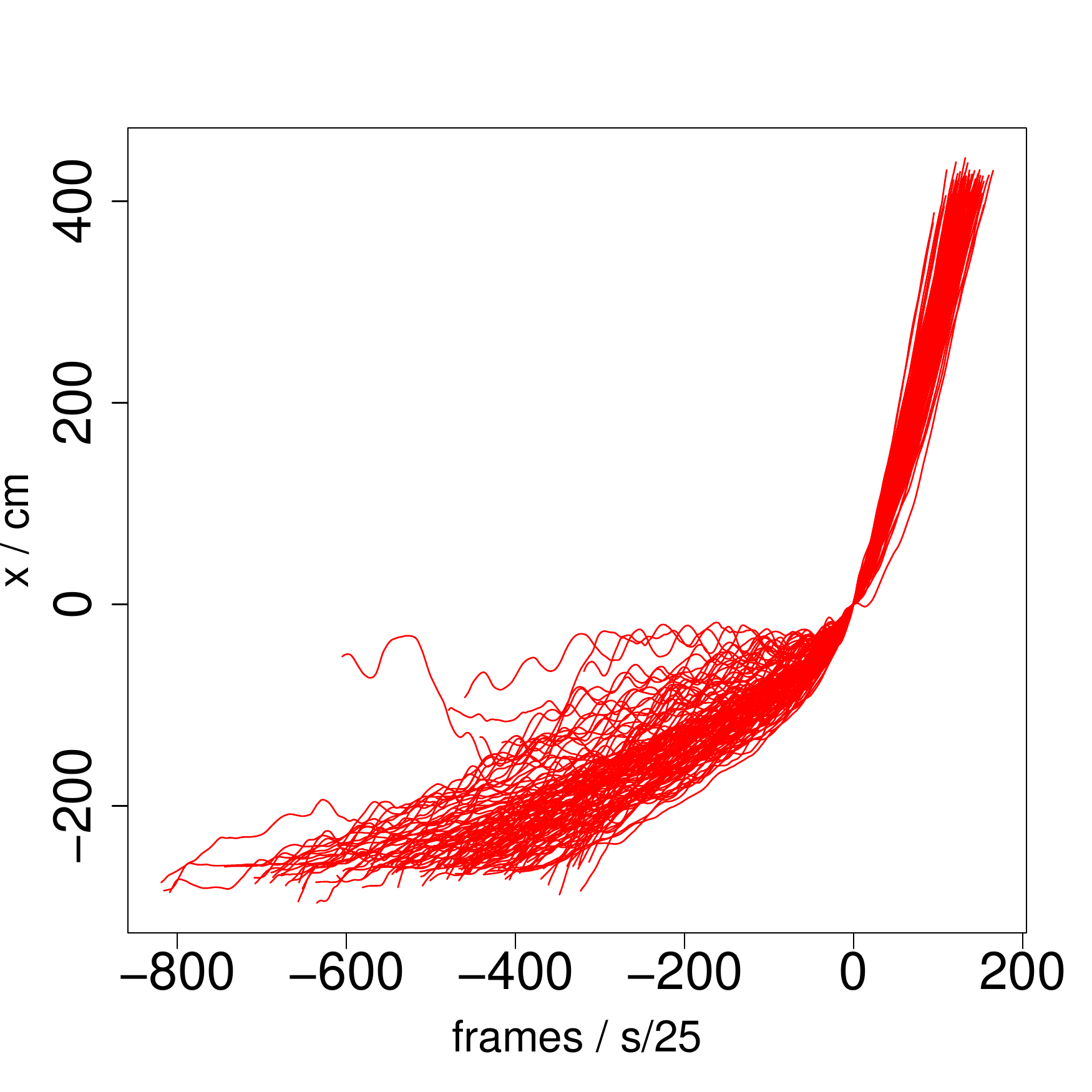}
  \includegraphics[width=0.32\textwidth]{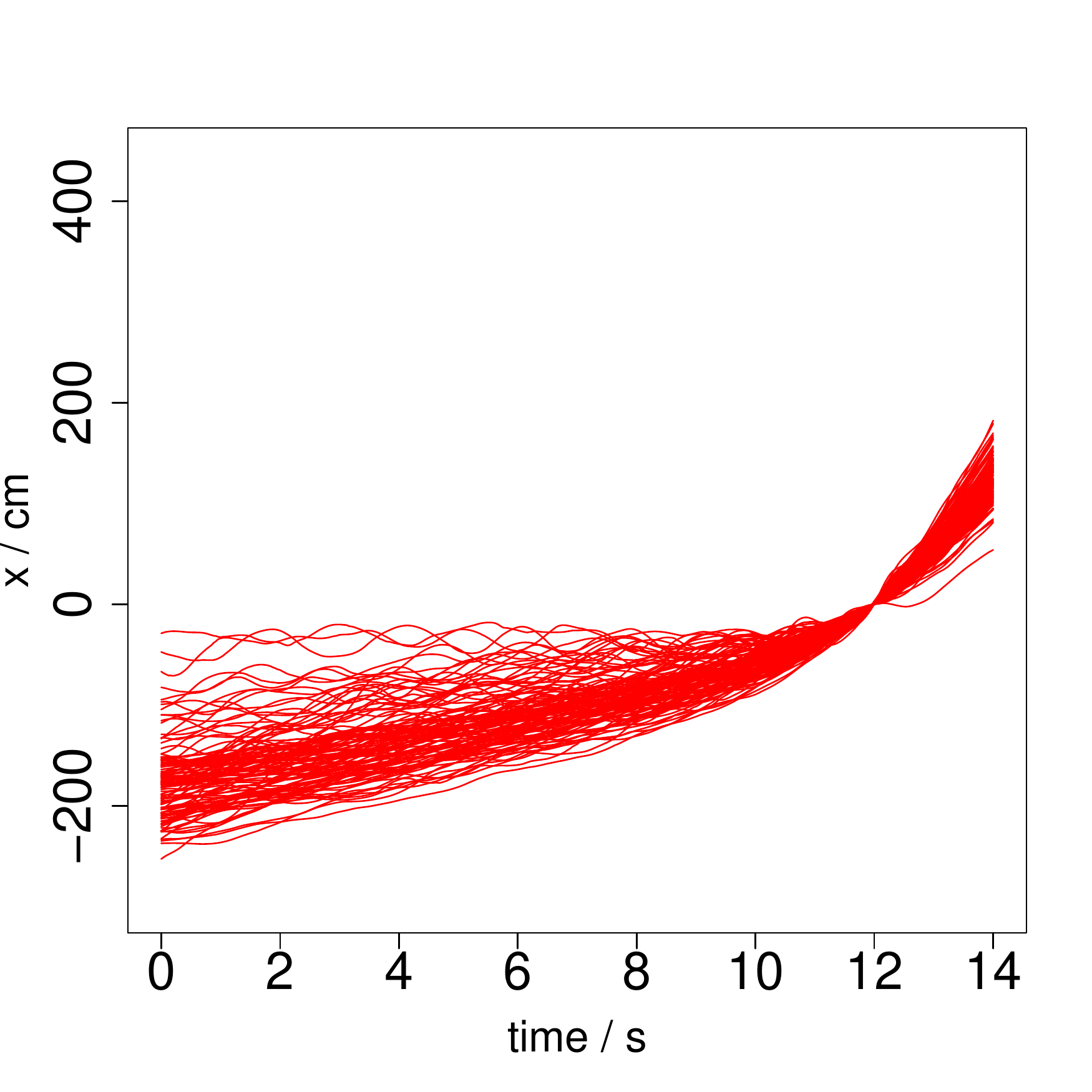}

  \caption{Plots of $x$-coordinates of pedestrians (experimental data): raw data (left), stationary data (middle) and with individual time in the interval [0,14] s (right). One Frame in the first two panels corresponds to 1/25 s.}\label{fig:formatting}
\end{figure}

Furthermore, the considered time interval has to be the same for all agents.
Therefore we account only to the pedestrians, who need more than 12 seconds to reach the exit.
Furthermore, the trajectory is tracked only two seconds after the passage of the bottleneck entrance.
The analysis of the experimental data and the models thus is based on trajectories in individual time intervals that range from $-12$\,s before passage through the bottleneck by the individual to $+2$\,s afterwards.
This makes a total time of $14$\,s.
In order to avoid negative time values, we start each pedestrian trajectory at time $t=0$\,s such that passage through the door for each individual  occurs at $t=12$\,s, exactly.
From now on we work with this time scale.
Figure~\ref{fig:formatting} visualizes the formatting steps.
After reformatting, a total of 118 pedestrian trajectories were available for the experiment, 121 for the SFM and also 121 for the GCFM model, respectively.

\begin{figure}[htb]
  \centering
  \includegraphics[width=0.5\textwidth]{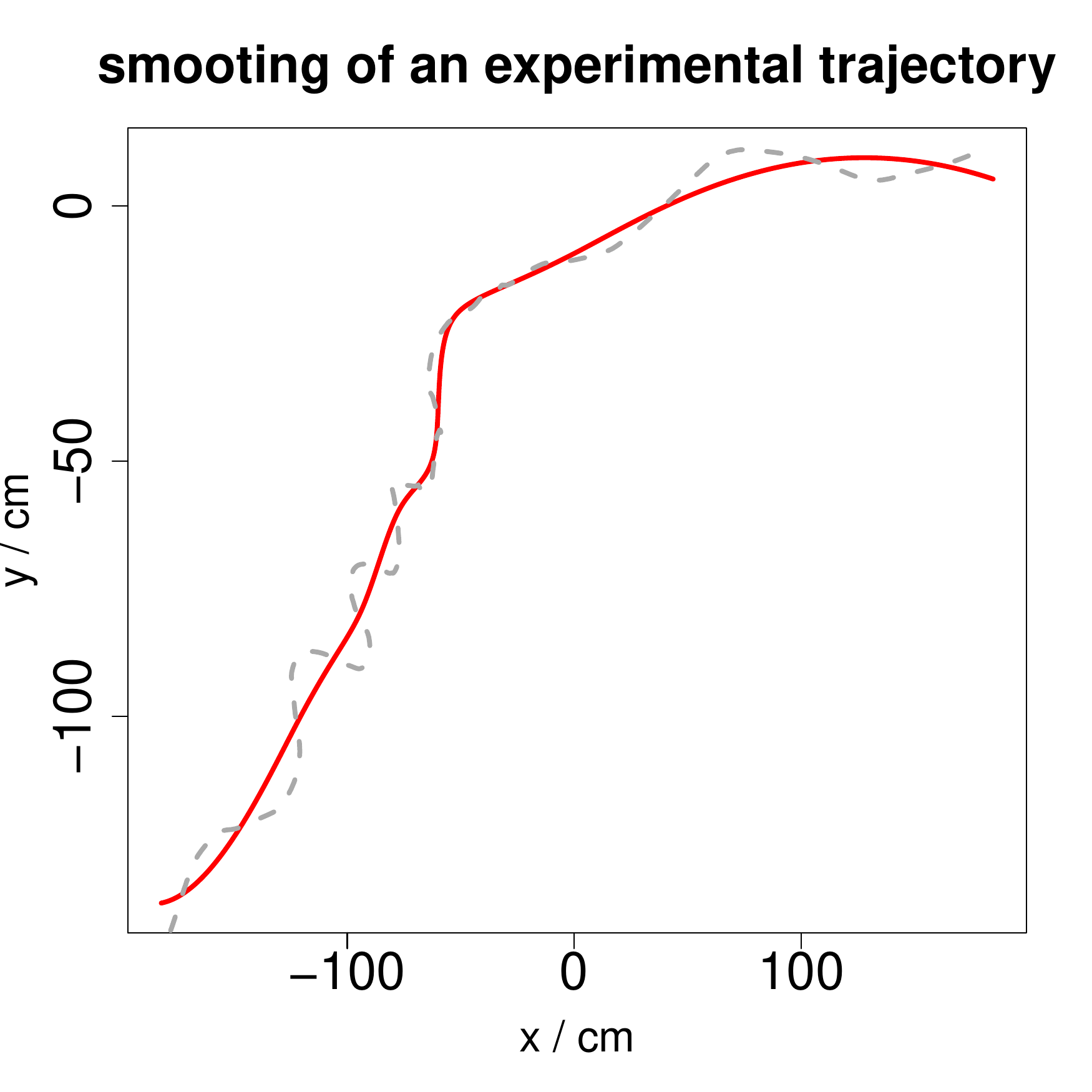}
  
  \caption{Smoothing of a experimental trajectory (dotted blue) with a B-spline basis of dimension 10 (solid red).}\label{fig:smooth}
\end{figure}

The experimental data contain the swaying caused by the bipedal locomotion of pedestrians in combination with the tracking of markers on the head.
But the SFM as well as the GCFM model the movement of the centre of mass neglecting the bipedal locomotion, which produce swaying-free trajectories.
Therefore, we smooth the data before the analysis in order to filter out the lateral swaying.
It turns out that the regression with a B-spline basis containing 10 elements with nodes equally distributed over the underlying time interval $[0,14]$ effectively removes swaying while properly reproducing the other features of the individual's trajectories, see Figure~\ref{fig:smooth}.

\subsection{\label{sec:average}Average trajectories}
As PCA components describe variation around some mean value, it is essential to analyse average functions $\bar x(t)$ and $\bar y(t)$.
\begin{figure}[htb]
  \centering
  \includegraphics[width=0.32\textwidth]{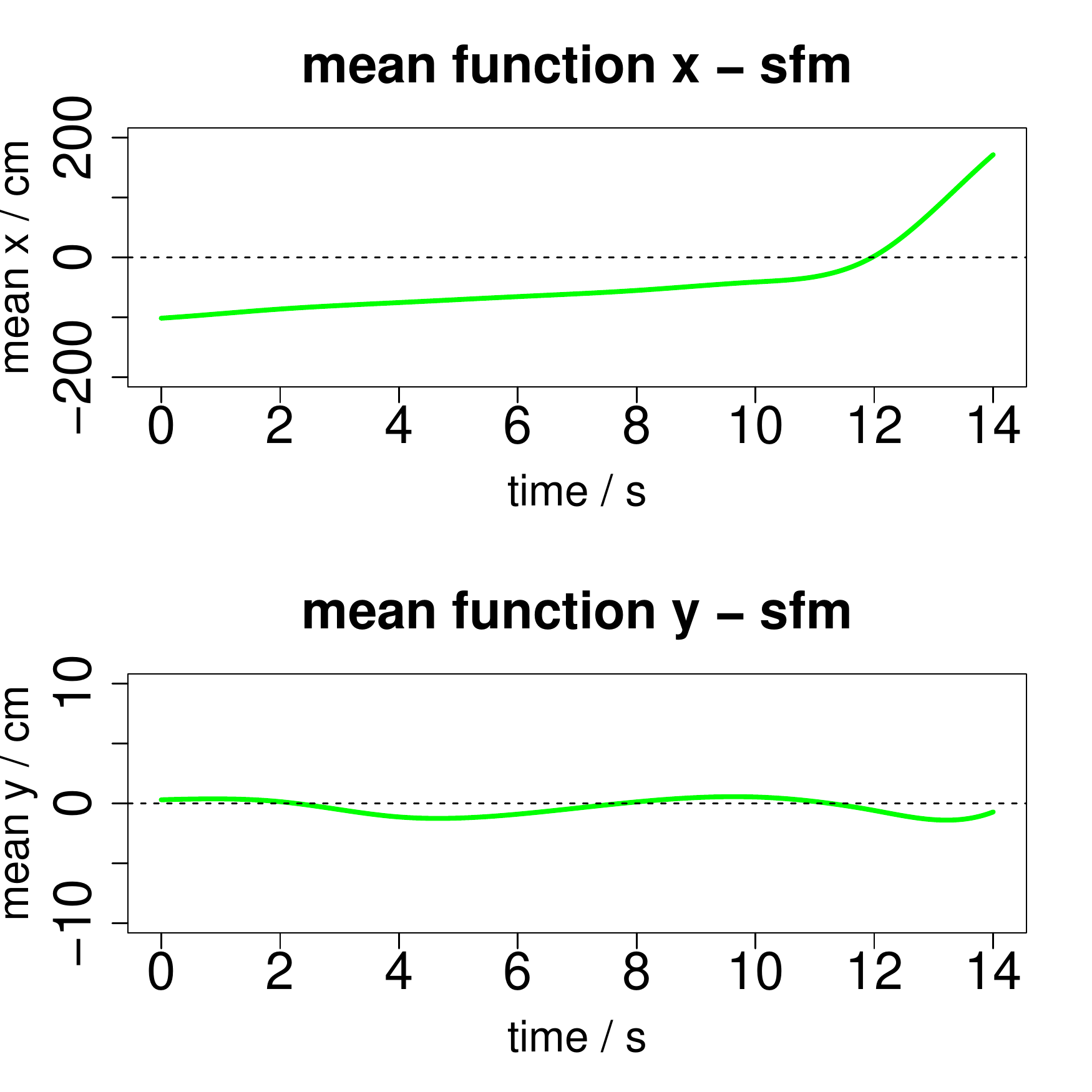}
  \includegraphics[width=0.32\textwidth]{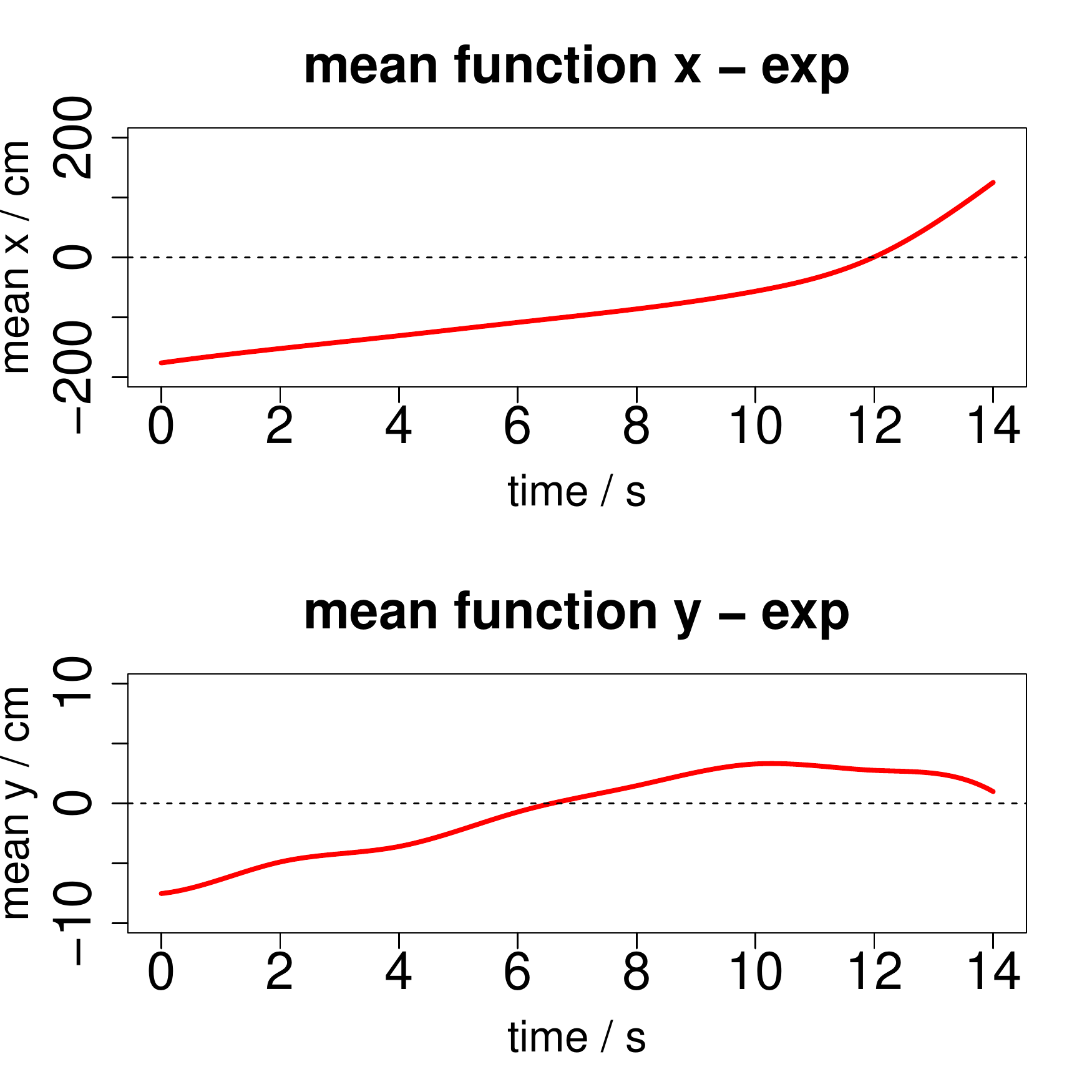}
  \includegraphics[width=0.32\textwidth]{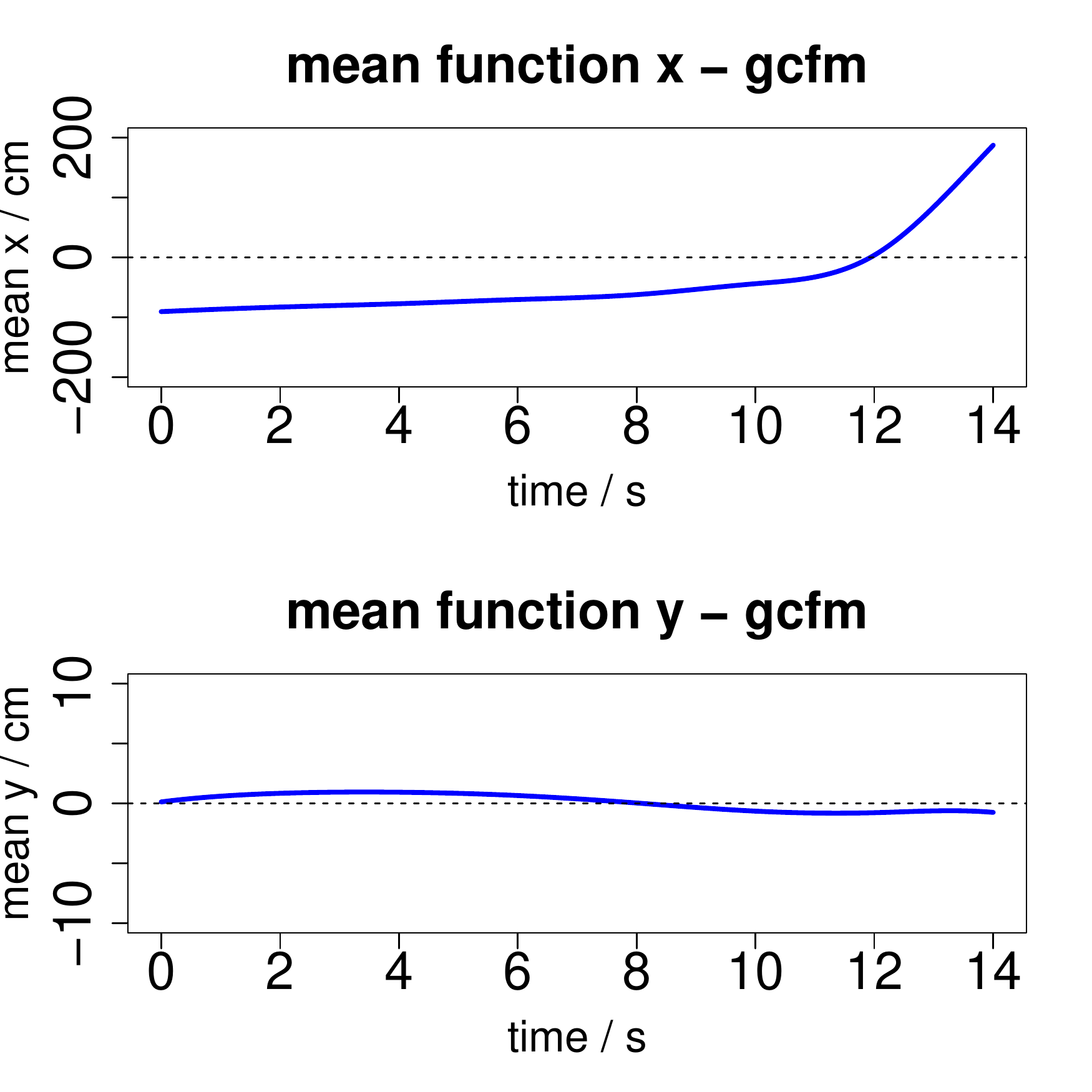}
  \caption{Average curves for position vs time for the SFM (left), experiment (middle) and GCFM (right): $x$-ccordinate (top) and $y$-coordinate (right).}\label{fig:averageXY}
\end{figure}

Figure~\ref{fig:averageXY} shows the mean functions $\bar x(t)$ and $\bar y(t)$ of the $x$- and $y$-components for the experiment and the models.

When we examine the $x$-component, we identify clearly  discrepancies between the experiment and both models.
The average pedestrian in the experiment shows a nearly linear progress to reach the exit.
The acceleration after passing the bottleneck, i.e.\ the increase in the slope, is modest.
In contrast, both SFM and GCFM show a slower progress of the average pedestrian through the crowd and a much more pronounced acceleration after the passage of the bottleneck.
While the latter deviation from experiment is about the same for both models, the underestimation in the slope from the experimentally observed value is bigger for the GCFM model.
Thus both models overestimate the dwell time indicating a missing anticipation and cooperation of the modelled pedestrians.
However the SFM produces this effect  by a lesser amount.

The $y$-component is described by both models in a satisfactory manner, as trend lines only move at a scale of a few centimeters from the center of the bottleneck.
At least for the simulated data this is due to the left hand - right hand reflection symmetry of the agents in both models and the (approximate) symmetry of initial positions with respect to the $y=0$ axis, i.e.\ the center line through the bottleneck.
In the experiment, a certain asymmetric behavior is visible for the mean $y$-position of the trajectories over time.
In average, the pedestrians approach the bottleneck coming slightly from the left seen from the direction of progress.
Interestingly, this asymmetry can not be traced back to the initial conditions, as these are the same for the experimental and the simulated trajectories.

This behavior is absent in both models, which have left-right symmetries in their respective constituting equations.

\subsection{\label{sec:pcastrength}PCA eigenvalues}
Having analyzed the average behavior, we now turn to the question, how well the models describe fluctuations in pedestrian data around the averages.
We start with the absolute strength of PCA variability, which is represented by the PCA eigenvalues $\lambda_i$, $i=1,\ldots,10$, as we are using a 10 dimensional spline basis.
At the same time, we also consider the cumulative relative strength $\varrho_j$ in order to measure the concentration or dispersion of variability in experimental or the simulated data.
\begin{figure}[htb]
  \centering
  \includegraphics[width=0.49\textwidth]{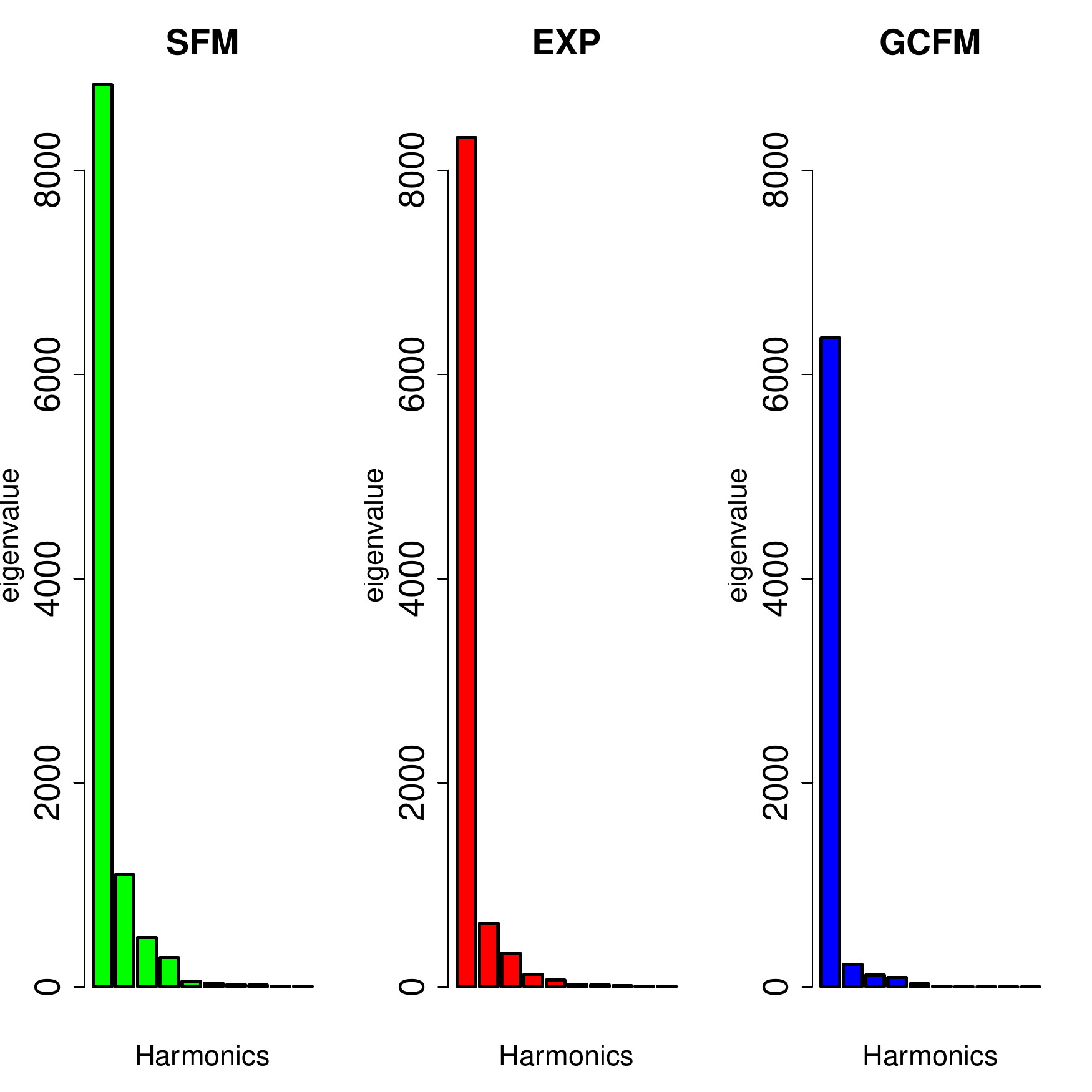}
  \includegraphics[width=0.49\textwidth]{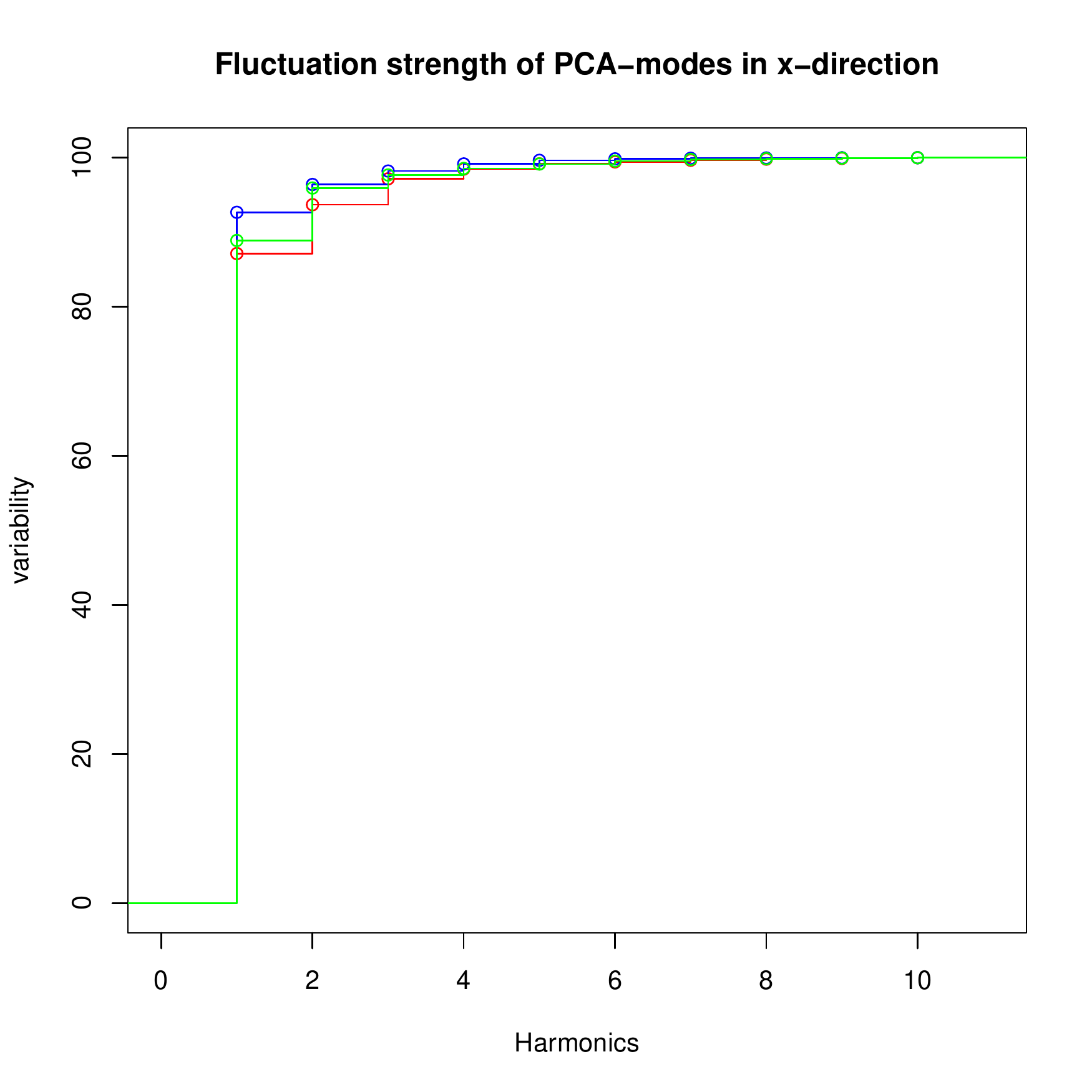}
  \caption{Left: Barplot of absolute fluctuation strength (eigenvalues) of the 10 PCA-modes (harmonics) for $x$-position over time for the SFM-model, Experiment and GCFM-model. Right: Cumulative relative strength of PCA-modes over all 10 harmonics.}\label{fig:strengthX}
\end{figure}
We first consider PCA mode strength for $x$-position over time as given in Figure~\ref{fig:strengthX}.
These modes describe typical patterns of pedestrians lagging behind and being in front of the trajectory of the average pedestrian.
The SFM overestimates the total amount of statistical deviation in $x$-position from the average $x$-position by  approximately 12\% as compared with the experiment. Also, concentration of variability in the first mode is slightly lower than in the experiment.
In the GCFM, the total level of $x$-position variation is  underestimated by 37\% of the total variation.
The relative concentration in the fist mode is higher as in the experimental data by an amount comparable to the SFM, but in the opposite direction.
For the values of total variation and Gini indices confer Table~\ref{tab:TotalGini}.
\begin{figure}[htb]
  \centering
  \includegraphics[width=0.49\textwidth]{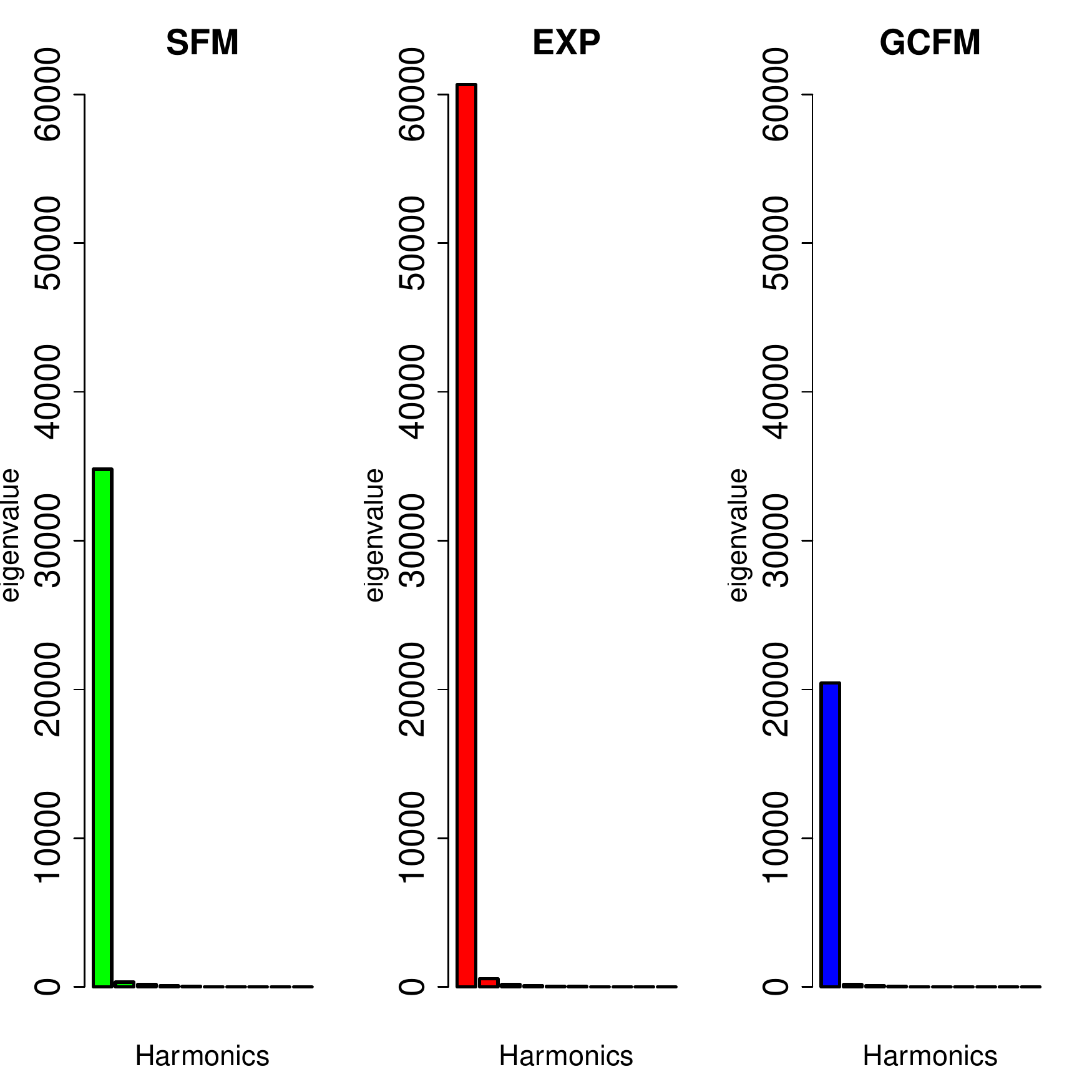}
  \includegraphics[width=0.49\textwidth]{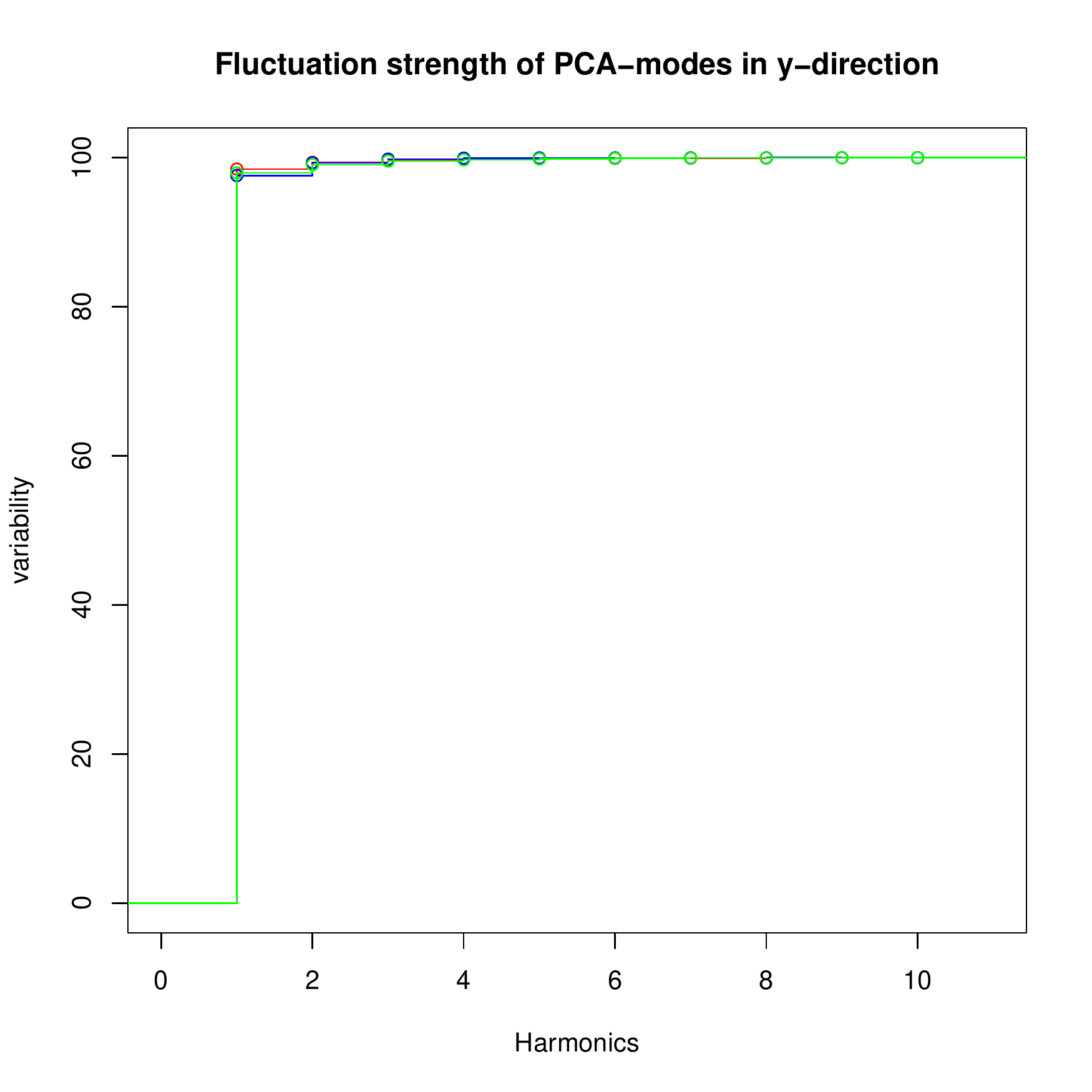}
  
  \caption{Left: Barplot of absolute fluctuation strength (eigenvalues) of the 10 PCA-modes (harmonics) for $y$-position over time for the SFM-model, Experiment and GCFM-model. Right: Cumulative relative strength of PCA-modes over all 10 harmonics.}\label{fig:strengthY}
\end{figure}

In Figure~\ref{fig:strengthY} the PCA-modes for the statistical $y$-fluctuation around the average $y$-position (essentially $y=0$) is displayed.
The experiment and both simulations all show that basically only one mode is active representing the axially symmetric  shape of the jammed area in front of the bottleneck.
The size of this area is underestimated by both models.
The GCFM predicts a pronouncedly reduced area in the $y$-direction covered by trajectories passing the bottleneck in the next 12 seconds, showing a total $y$-variation of  $\approx 1/3 (33.6\%)$ of the experimental data.
The same figure of underestimation of $y$-variation for the SFM compared with experiment is 57\%.
Total variations and Gini indices can again be found in Table~\ref{tab:TotalGini}.

\begin{table}[h]
  \centering
  {
    
    \begin{tabular}{|r|r|r|r|}
      \hline
      \textbf{Tot.Var}& SFM & Experiment & GCFM \\
      \hline
      $x$-position & 10863& 9550& 6835\\
      $y$-position & 35469& 61619& 20749 \\
      \hline
    \end{tabular}
    ~~
    \begin{tabular}{|r|r|r|r|}
      \hline
      \textbf{Gini}& SFM & Experiment & GCFM \\
      \hline
      $x$-position & 0.83 & 0.85 & 0.87 \\
      $y$-position & 0.89 & 0.89 & 0.89 \\
      \hline
    \end{tabular}
  }
  \caption{Total variation (left) and Gini indices (right) for the eigenvalues of the PCA.}
  \label{tab:TotalGini}
\end{table}
\subsection{\label{sec:pcaxy}PCA modes}

\begin{figure}[htb]
  \centering
  \includegraphics[width=0.32\textwidth]{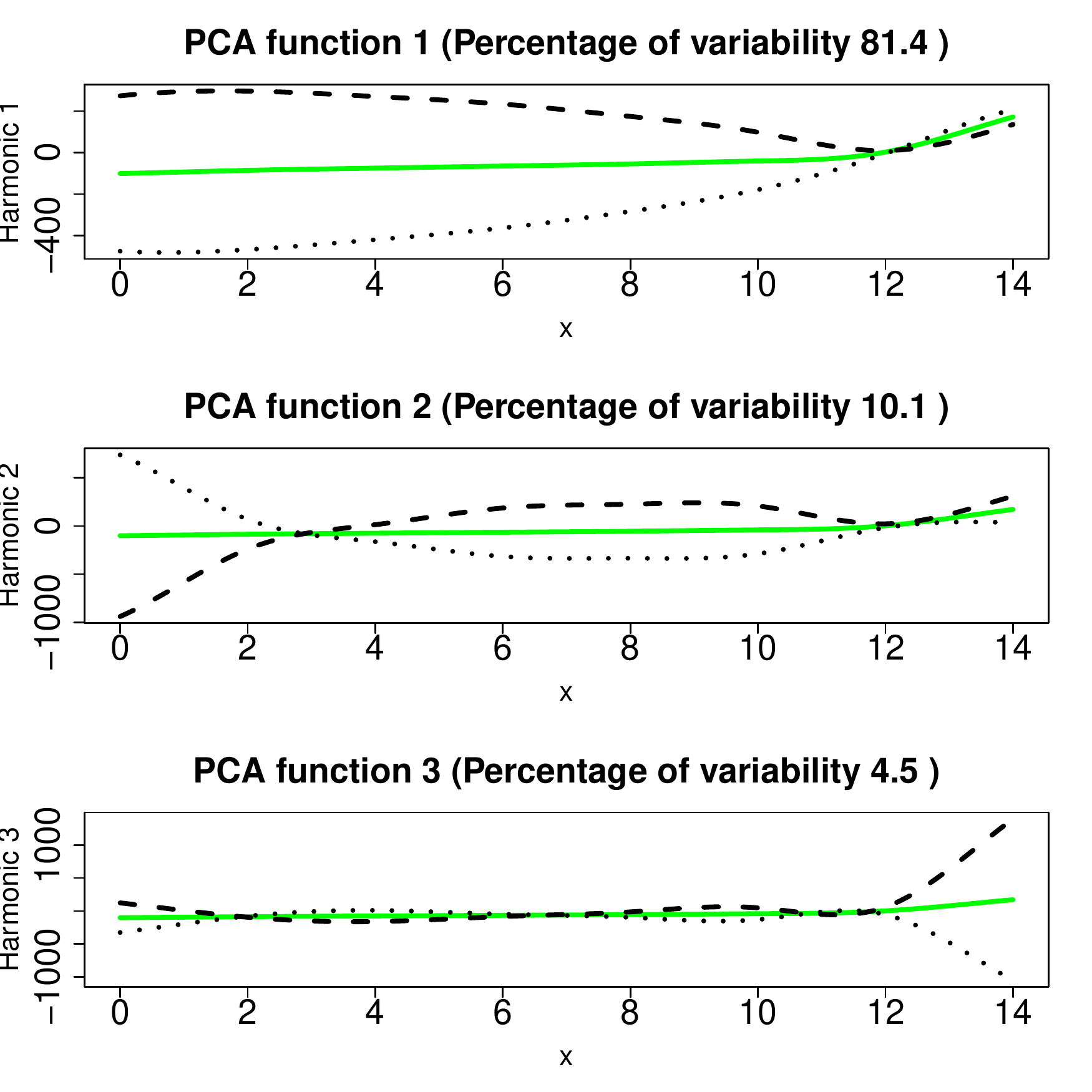}
  \includegraphics[width=0.32\textwidth]{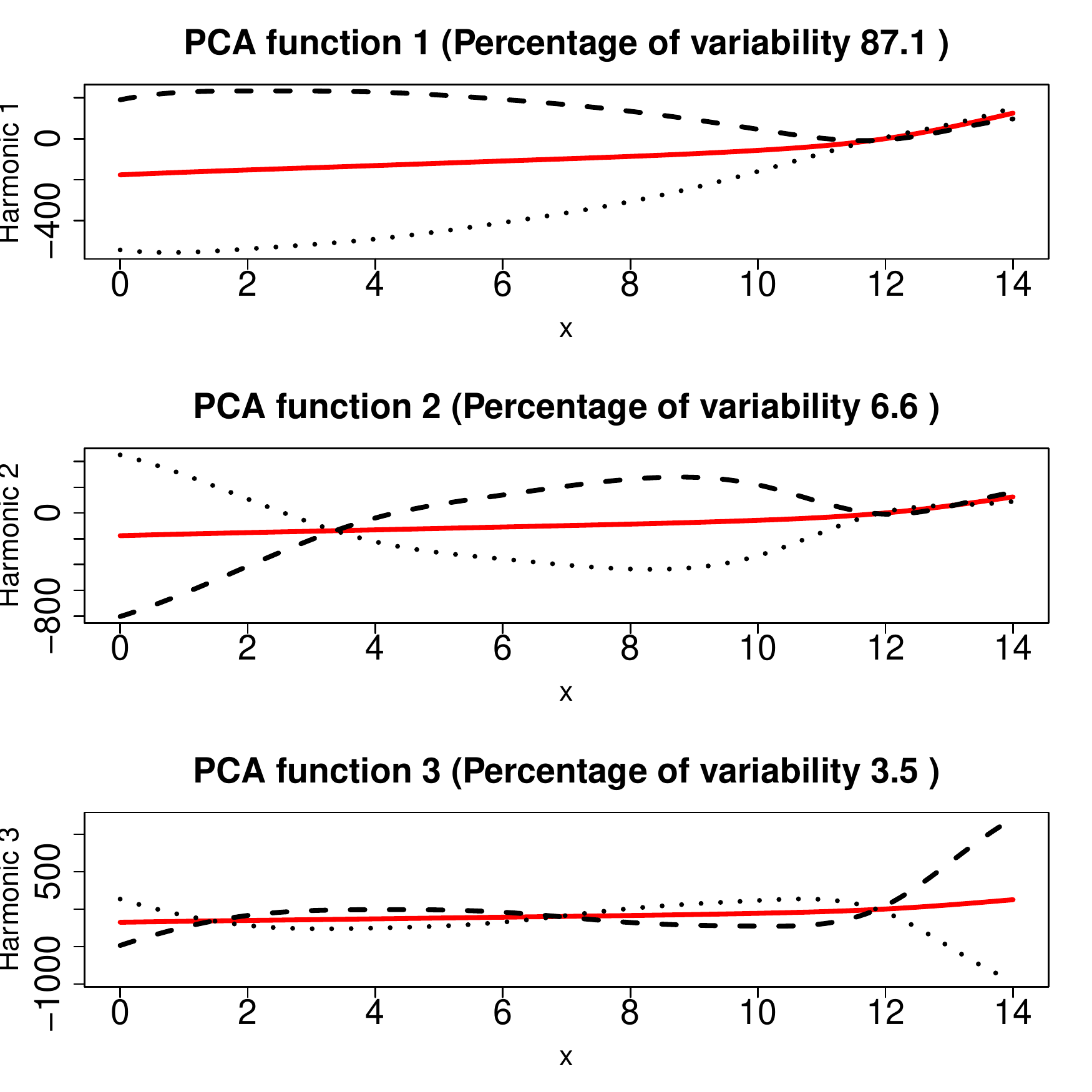}
  \includegraphics[width=0.32\textwidth]{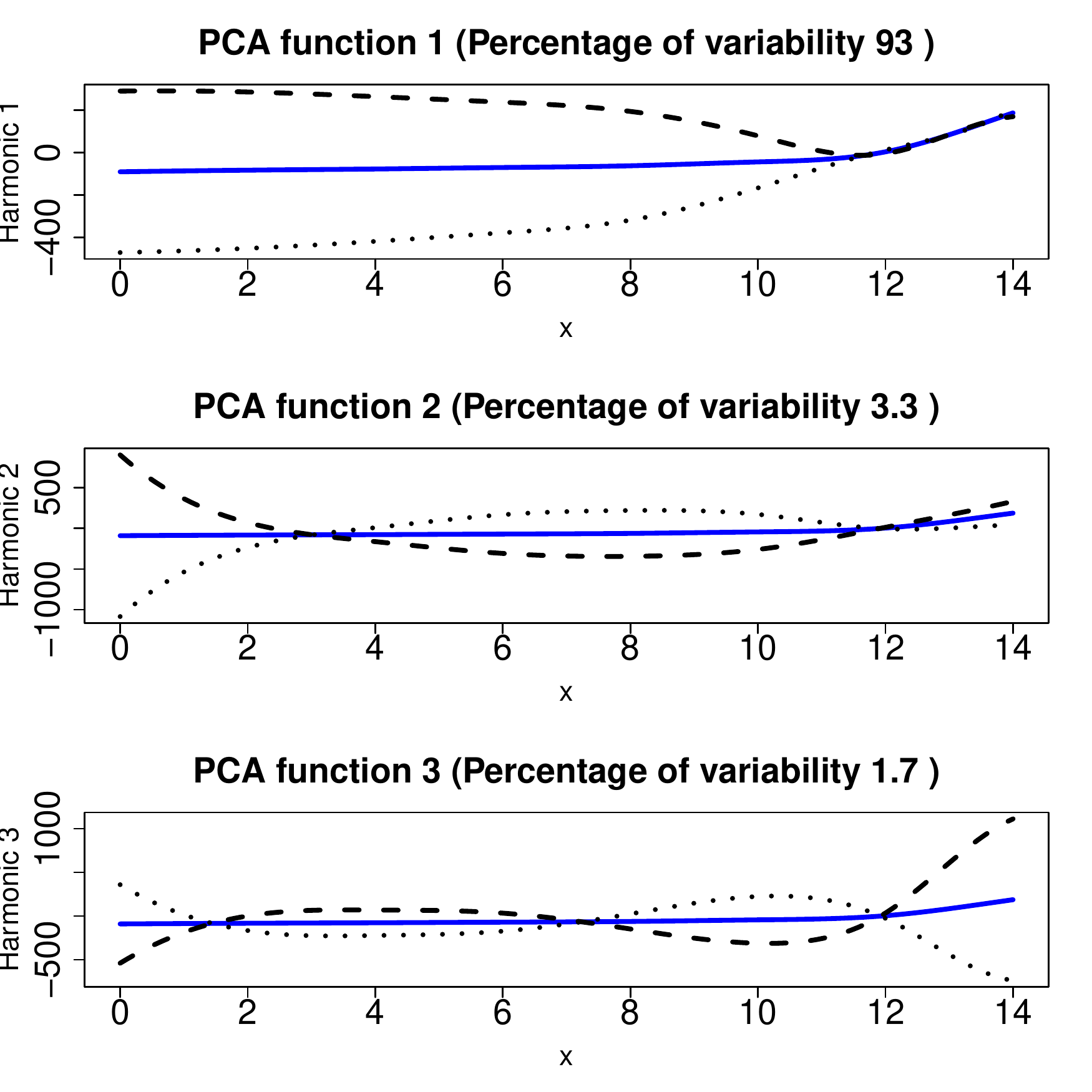}
  
  \caption{PCA components in for $x$-position over time  for the SFM model (left), experiment (middle) and GCFM model (right). The first three harmonics are displayed.}\label{fig:PCAx}
\end{figure}

Figure~\ref{fig:PCAx} shows the principal fluctuation components of  the $x$-position for the experiment and the models.
From the aforementioned eigenvalue analysis (Figure~\ref{fig:strengthX}) we can conclude that fluctuations can be mainly described by the first three principal components.

Let us now give an interpretation to the morphology of the PCA components.
The first principal component describes the tolerance between the initial positions of the pedestrians 12 seconds before passing the bottleneck.
Two pedestrians reach the bottleneck at the same reference time $t=12$.
Due to the fact that some pedestrian starts with a higher or lower $x$-distance to the bottleneck than the the others, at $t=0$, a statistical variation in $x$-positions occurs.
This could be called ``slipping-through-effect'' because faster pedestrians find more favorable configurations of fellow pedestrians ahead which allows a faster passage through the crowd.

The second and third principal component describe an effect which we can associate to long stop and go behavior in different lanes in a traffic jam:
One trajectory is temporary faster than the other, but afterwards it is the other way round.
In the case of the experiment and the GCFM, the third principal component also shows different velocity patterns after the bottleneck.

The morphological comparison of the experimental data with the SFM and GCFM  shows that the points of intersection of the first two principal components are nearly at
the same times.
Thus both models reproduce the qualitative behavior of statistical fluctuations in the pedestrians $x$-positions over time quite well.
The main difference thus lies in the different activation strength of the ``slipping-through'' mode.

\begin{figure}[htb]
  \centering
  \includegraphics[width=0.32\textwidth]{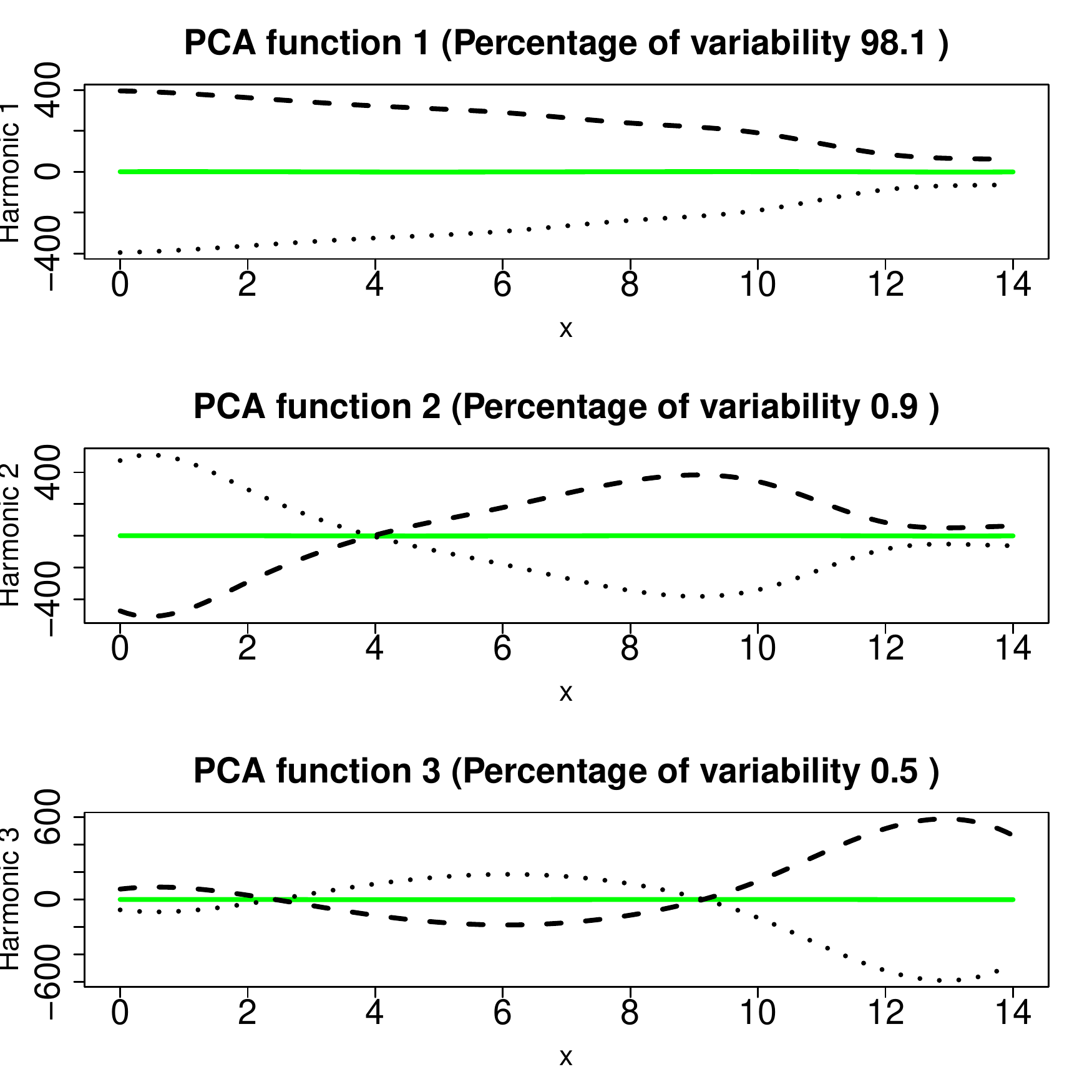}
  \includegraphics[width=0.32\textwidth]{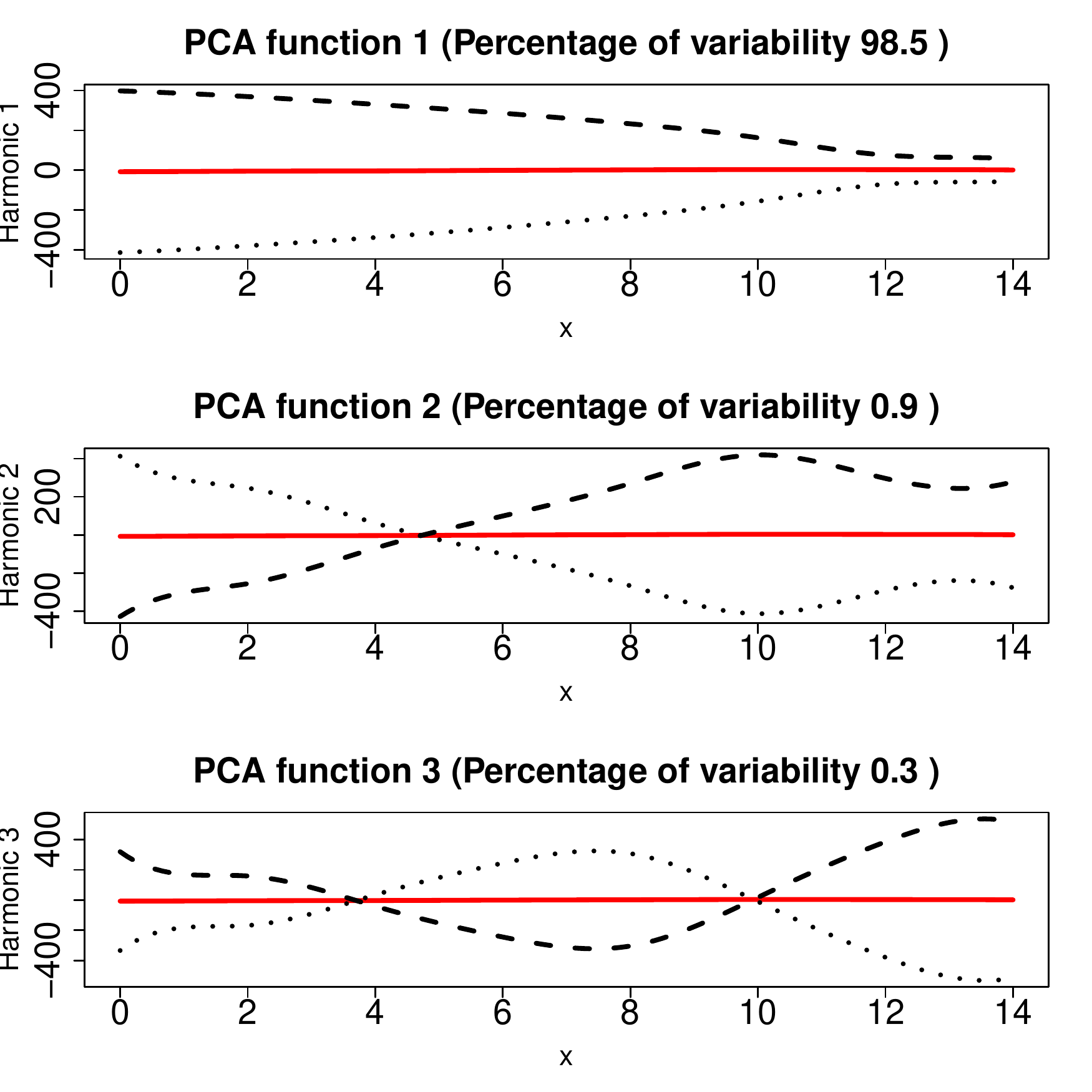}
  \includegraphics[width=0.32\textwidth]{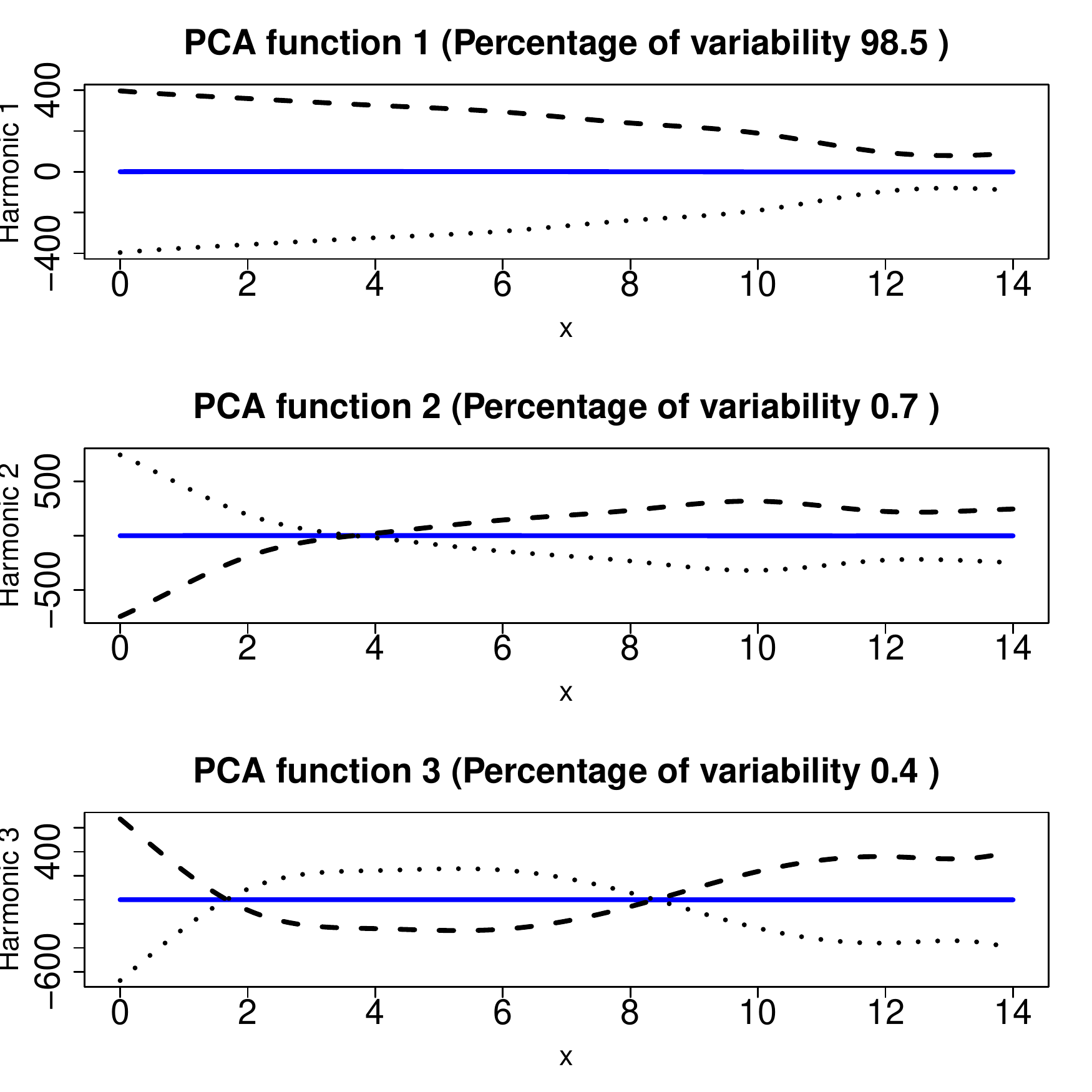}
  
  \caption{PCA components in the y direction for the SFM model (left), experiment (middle) and GCFM model (right). The first three harmonics are displayed.}
  
  \label{fig:PCAy}
\end{figure}

Figure~\ref{fig:PCAy} shows the PCA of  the $y$-components for the experiment and the models.
We observe that the variability of the data  can be mainly described by the first principal component which represents the shape of the crowd in front of the door.
The first principal component of both models describe the experiential data acceptably well.
Also the higher modes are of quite similar shape, although they should be neglected since they  hardly contribute to the total variation.

\subsection{\label{sec:DeviationResults} Evaluation of deviation measures}

Lastly in this section, we want to compare the deviation measures of the respective simulation model with the experiment.
The results are summarised in~\ref{tab:HSnorm}.

\begin{table}[h]
  \centering
  {
    \begin{tabular}{|r|r|r|}
      \hline
      {\bf $L^2$-norm} & EXP-GCFM & EXP-SFM \\
      \hline
      $x$-position & 169.8 &  166.0 \\
      $y$-position & 15.34 &  13.20 \\
      \hline
    \end{tabular}
    ~~
    \begin{tabular}{|r|r|r|}
      \hline
      {\bf HS-norm} & EXP-GCFM & EXP-SFM \\
      \hline
      $x$-position & 2291& 1282\\
      $y$-position & 4449 & 2880\\
      \hline
    \end{tabular}
  }
  \caption{Deviations between experimental data and models data using the $L^2$-norm (left) Hilbert-Schmidt norm (right).}
  \label{tab:HSnorm}
\end{table}

The average squared distance of $x$- and $y$- coordinates of the trajectories as function over time  is of the same order of magnitude for both models. 

A slight advantage can however be attributed to the SFM-model.
This effect is even more pronounced in the Hilbert-Schmidt norm that measures the distance to the experiment in the fluctuation structure of measured and simulated data.
\section{\label{sec:StatInf}Statistical inference based on the bootstrap}
In the previous section, we evaluated the total variation and the Gini coefficient or deviation measures for the average behavior and the fluctuation structure ($L^2$- and HS-norms, respectively) in order to compare simulated and experimental data.
This descriptive approach however leaves open the question, to which extent these findings depend on the intrinsic stochastic nature of pedestrian trajectories and to
which extent they are due to structural differences between  simulated  agents in the models  and  real pedestrians observed in the experiment.
In the present section, we describe and apply a simulation-based test procedure in order to clarify, to what extent the observed differences between models and experiment are statistically significant.

\subsection{\label{sec:BootMeth}Bootstrapping PCA scores}
\begin{figure}[htb]
  \centering
  \includegraphics[width=0.49\textwidth]{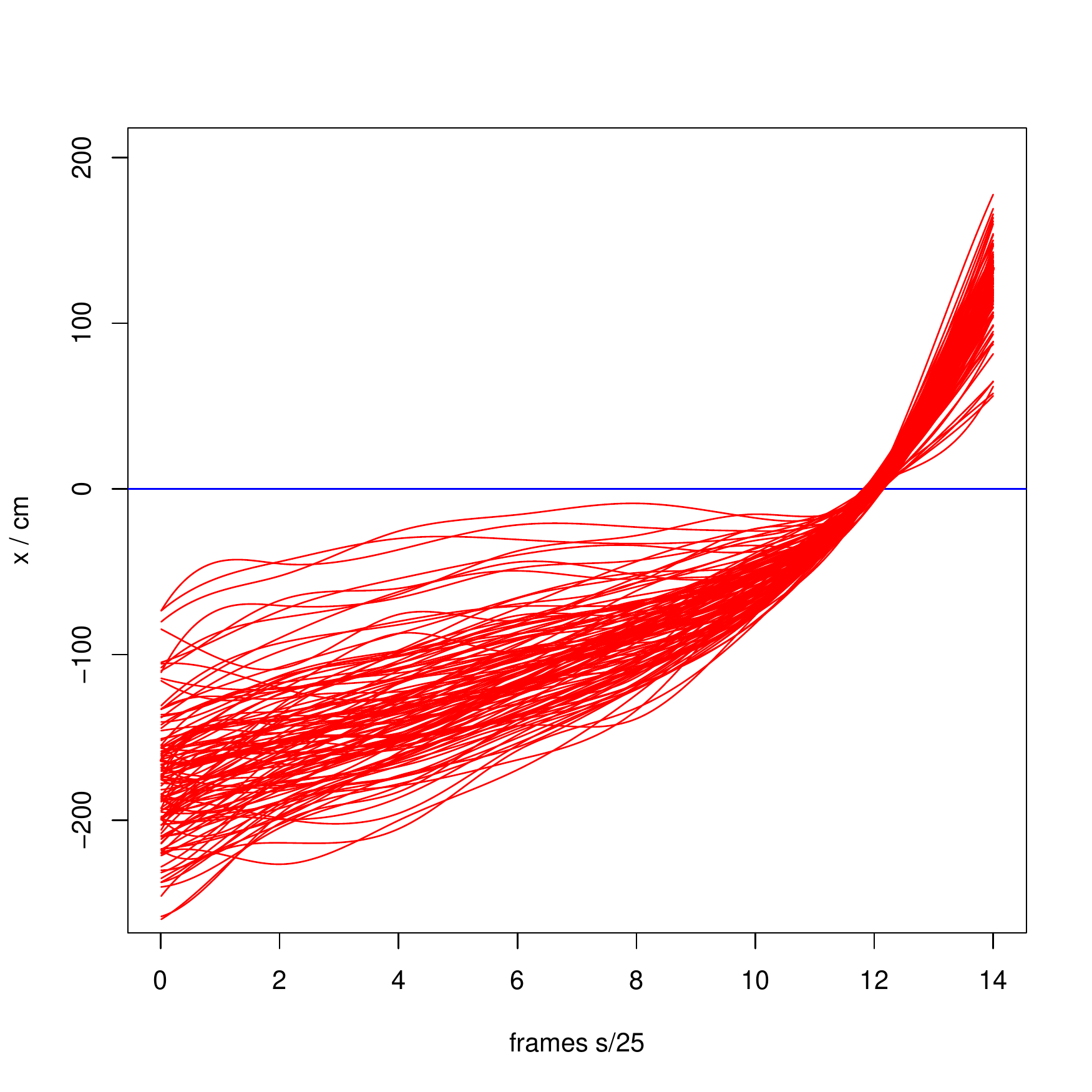}
  \includegraphics[width=0.49\textwidth]{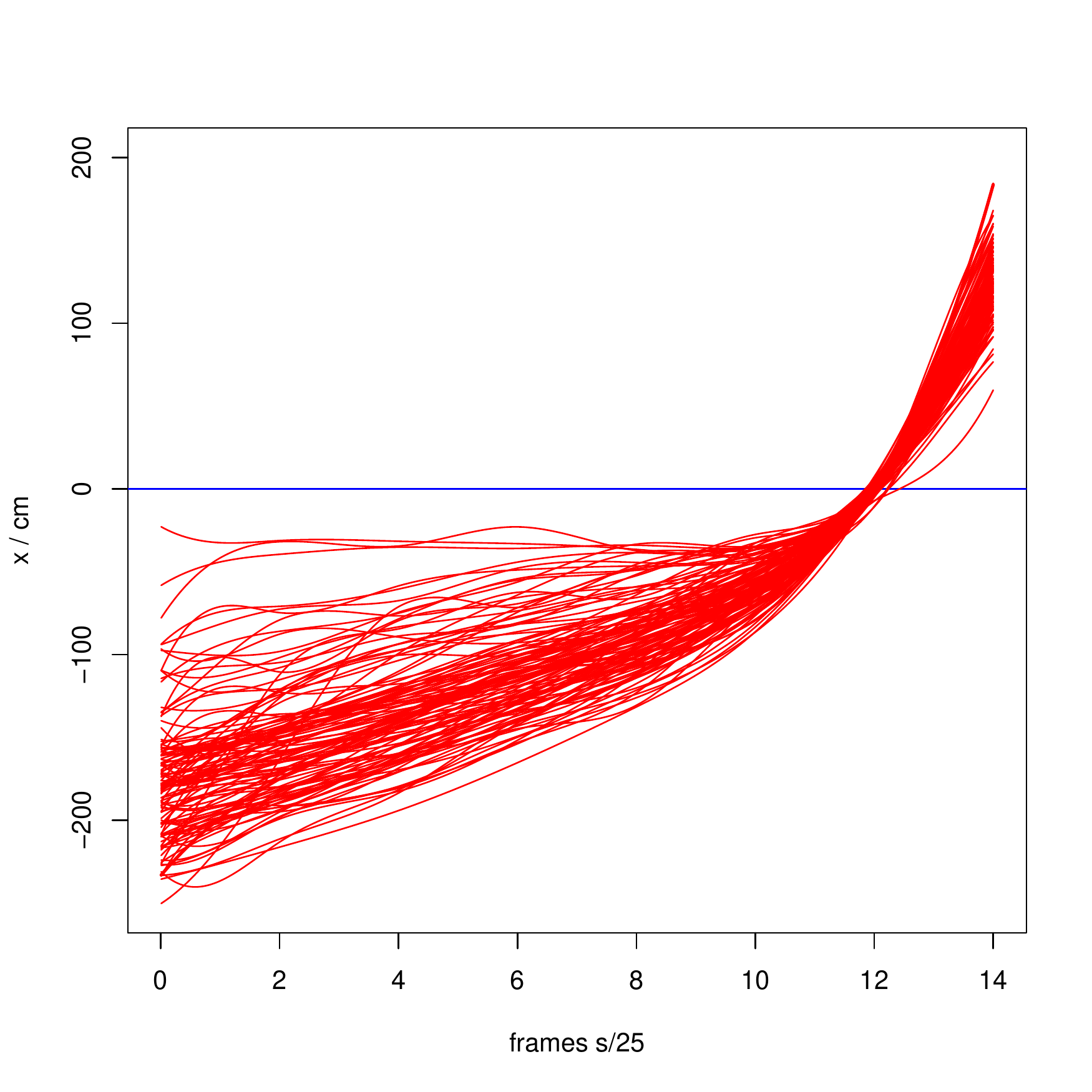}
  
  \caption{Plots of $x$-coordinates of pedestrians by bootstrapped data (left) and experimental data (right).}
  \label{fig:Xweg_boot}
\end{figure}

As the basis of our statistical testing procedure, we use the bootstrap over the matrix of principal components from~\cite{DiaconisEfron1983,FisherCaffoSchwartzZipunnikov2014}.
We now shortly describe the bootstrap approach.
Given, e.g., the $i$-th $x$-value of the trajectory over time, $x_i(t)$, the score of this trajectory with respect to the principal component $\xi_j(t)$ is
\begin{align}
  \label{eqa:scores}
  s_{ij}^{(x)}&=\langle x_i(t)-\bar x(t),\xi_j(t)\rangle\nonumber \\
              &=\left\langle \sum_{l=1}^K\underbrace{(c_{i,l}-C_l)}_{{\bf \Delta}_{i,l}}\Phi_l(t),\sum_{k=1}^K{\bf b}_{jk}\Phi_k(t)\right\rangle\\
              &=({\bf \Delta}^T{\bf W}{\bf b}^T)_{i,j}~.\nonumber 
\end{align}
By construction, the scores $s_{i j}^{(x)}$ and $s_{i,j'}^{(x)}$ are (linearly) uncorrelated for $j\not=j'$.
Neglecting potential higher order correlations, we construct a virtual bootstrap sample from the scores of the experimental data by drawing with replacement, for $i,j$ fixed, $s_{i,j}^{(x,{\rm boot})}$ from the $N$ samples $s_{i,j}^{(x)}$ of the original scores with respect to the $j$-the principal component $\xi_j(t)$.
Doing this independently for $i=1,\ldots,N$ and $j=1,\ldots,K$ (remember, in our case $N=110$ is the number of experimental pedestrian trajectories and $K=10$ the number of principal components) we obtain the $N\times K$ bootstrap score matrix ${\bf s}^{(x,{\rm boot})}$.
The corresponding bootstrapped trajectories then are
\begin{align}
  \label{eqa:bootTrajectory}
  x_i^{({\rm boot} )}(t)&=[{\bf s}^{(x,{\rm boot})}\Phi(t)]_i+\bar x(t),~~i=1,\ldots N.
\end{align}
Figure~\ref{fig:Xweg_boot} shows the $x$-coordinates plots of pedestrian trajectories by bootstrapped and experimental data.

With this virtual data set, the PCA analysis is then repeated.
In particular, we obtain bootstrapped quantities for total variation and Gini index, as well as distance measures for the average behavior of the actual experiment and its virtual bootstrap replica.
This entire process is then repeated a sufficiently high number of times, such that $p$-values in the range of usual significance levels $\approx 1-5\%$ can safely be determined.
Here we generate  $10^4$ bootstrap samples, each containing $N=118$ virtual trajectories, and thereby obtain a simulated distribution for each of the aforementioned quantities.

The same iteration is repeated for the $y$-coordinate and the $x$-coordinate of the velocity $v_x$. 

For statistical testing, we generate two-sided confidence intervals for the total variation and the Gini coefficient and left open confidence intervals for the distance measures based on the empirical distributions of the respective quantities.
If the related quantities for the SFM and GCFM model are not contained in these confidence regions, we consider this as a positive test result for a deviation between experiment and model.

\subsection{\label{sec:GiniTest}Testing Gini indices and total variations}
\begin{figure}[htb]
  \centering
  \includegraphics[width=0.49\textwidth]{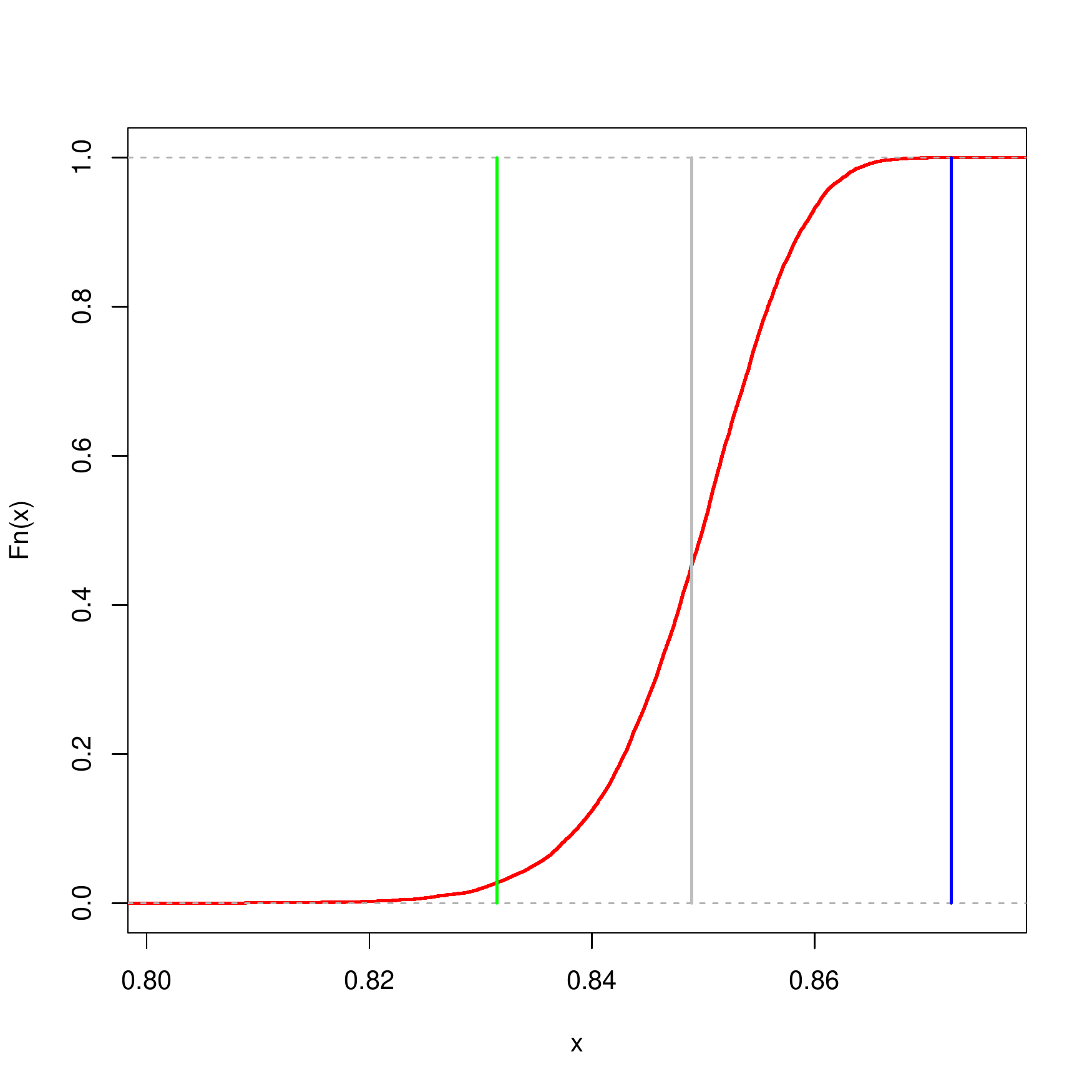}
  \includegraphics[width=0.49\textwidth]{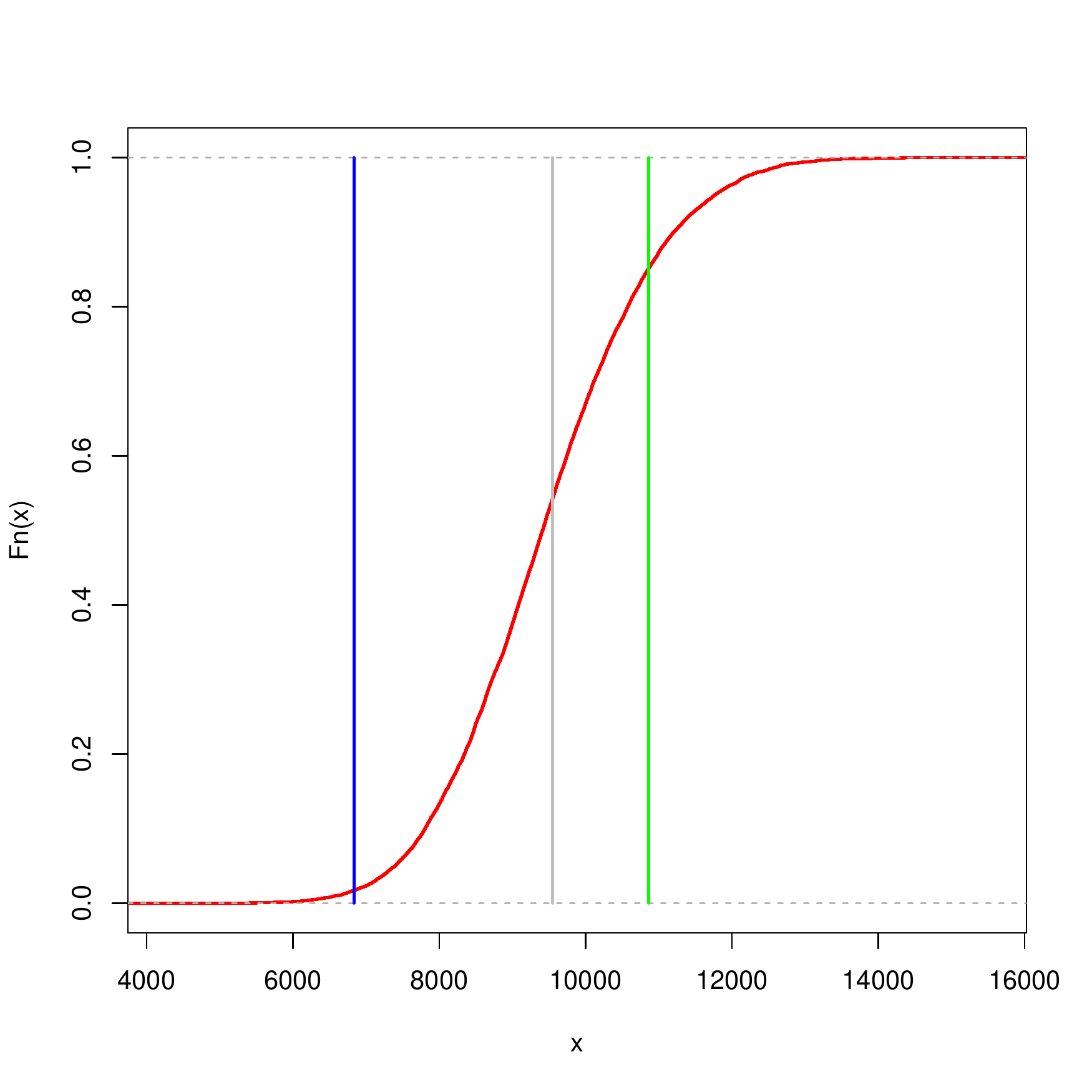}
  
  \caption{ECDF of Gini indices (left) and ECDF of total variation (right) from bootstrapped experimental data. Green, grey and blue vertical lines mark the values for the SFM, experiment and GCFM, respectively.}\label{fig:ECDF_GINI_total_EXP_boot}
\end{figure}

One of the advantages of using this bootstrap technique is to have the opportunity to examine the distributions of Gini indices and the total variations.
We compute for every bootstrap sample the Gini indices and total variations by the experimental data for $x$-coordinates, $y$-coordinates and $x$-velocities.
Afterwards, we are able to compute their empirical cumulative distribution functions (ECDF). Figure~\ref{fig:ECDF_GINI_total_EXP_boot} shows the ECDF of Gini
indices and total variations of $x$-coordinates by bootstrapped experimental data. The blue lines show the values of Gini index and total variation
by original experimental data.
The corresponding $p$-values, i.e.\ the critical level of statistical significance where the difference between model and experiment becomes  significant, are calculated on the basis of two-sided confidence regions of the bootstrapped distribution. The $p$-values are summarized in Table~\ref{tab:P_value_GINI_Tot}. Note that $p$-values below $10^{-3}$ become numerically unreliable for $10^4$ bootstrap samples and are set to zero.

\begin{table}[h]
  \centering
  {
    \begin{tabular}{|r|r|r|}
      \hline
      \multicolumn{3}{|c|}{\textbf{p-values for Gini index}} \\
      \hline
   & EXP/GCFM & EXP/SFM \\
      \hline
      $x$-position   & 0   &  0.055   \\
      $y$-position   & 0.610& 0.007\\
      \hline
    \end{tabular}
    ~~
    \begin{tabular}{|r|r|r|}
      \hline
      \multicolumn{3}{|c|}{\textbf{p-values for TotalVar}} \\
      \hline
   & EXP/GCFM & EXP/SFM \\
      \hline
      $x$-position   & 0.034  & 0.297 \\
      $y$-position   & 0      & 0     \\
      \hline
    \end{tabular}
  }
  \caption{The p-value of Gini index and total variations by bootstrapped experimental data.}
  \label{tab:P_value_GINI_Tot}
\end{table}

The $p$-values in Table~\ref{tab:P_value_GINI_Tot} show that statistical testing reveals significant differences between experiment and model with respect to several
Gini indices and total variations.
Only the Gini index of the GCFM and the total variation of the SFM do not produce significant differences (significance level $5\%$).

\subsection{\label{sec:L2HSTest}Tests based on deviation measures}
\begin{figure}[htb]
  \centering
  \includegraphics[width=0.49\textwidth]{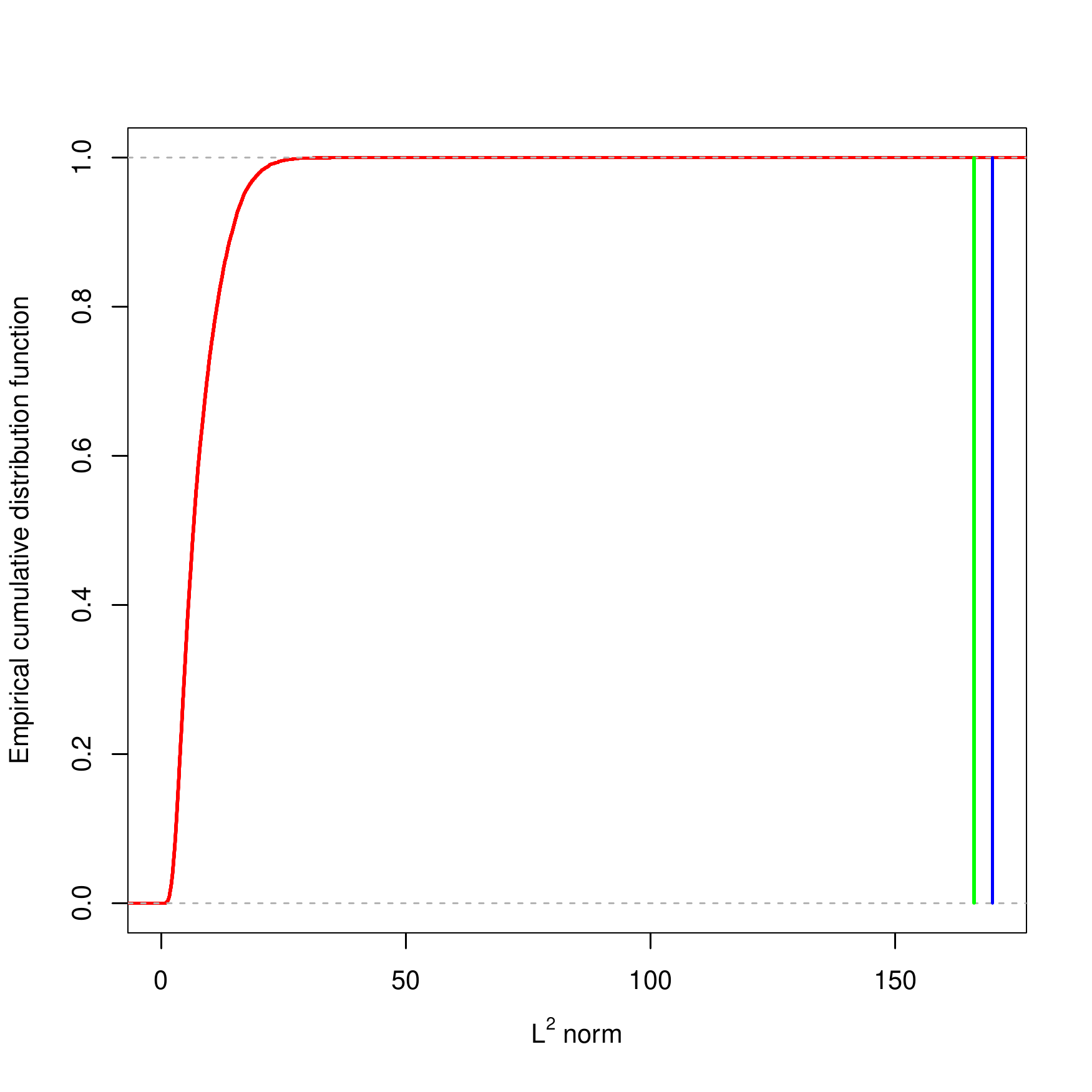}
  \includegraphics[width=0.49\textwidth]{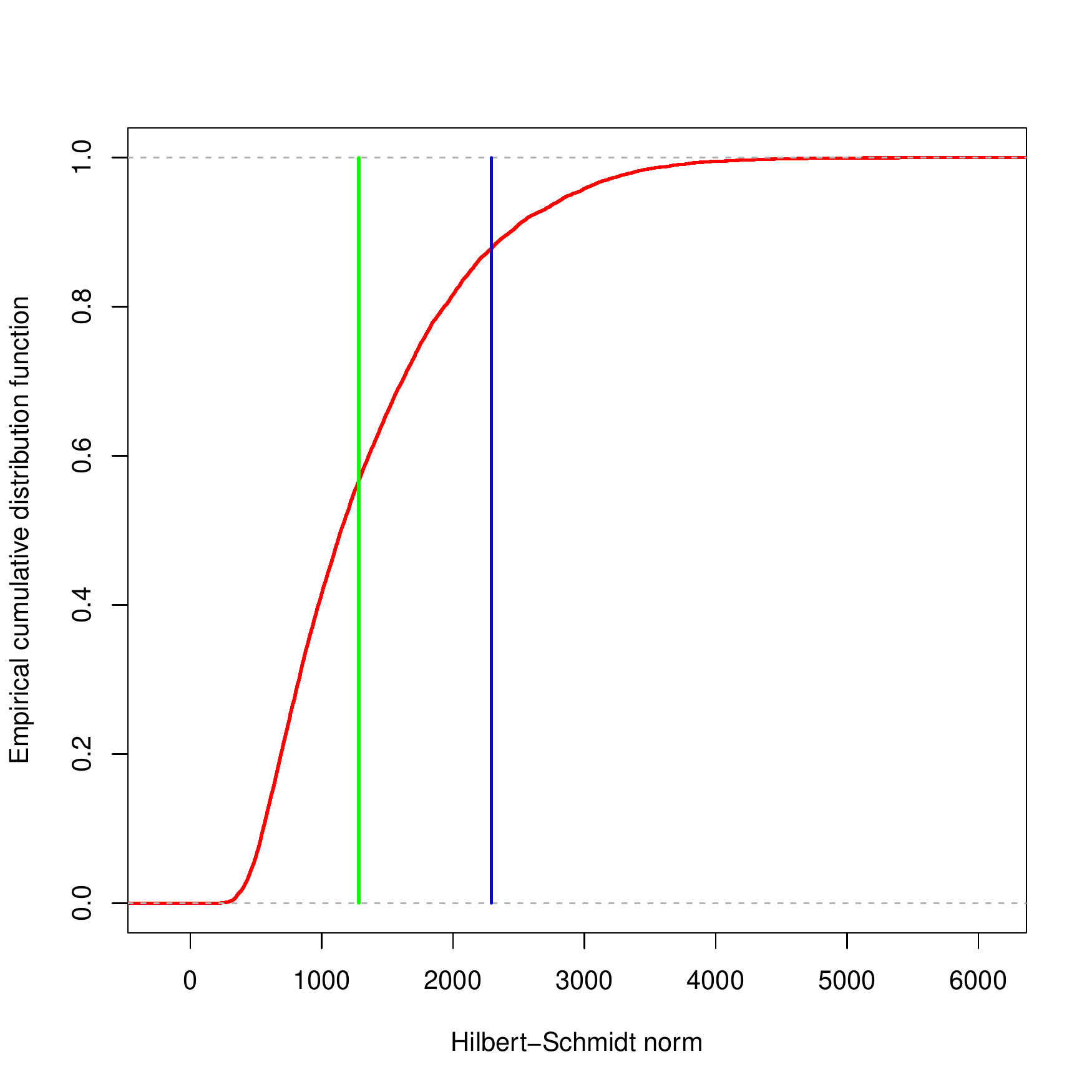}
  \caption{The ECDF of the $L^2$-norm difference of average $x(t)$ values (left) and the Hilbert Schmidt (HS) norm difference between covariance functions of $x(t)$ fluctuations of experimental and bootstrapped data. Green and blue vertical lines show $L^2$- and HS-norm distance between the experiment  SFM and GCFM models respectively.}\label{fig:HS_norm_EXP_GCFM_SFM_x}
\end{figure}

We are now interested to measure the deviations between bootstrapped and experimental data.
Firstly we compute for every bootstrap sample the $L^2$ norm of the difference between the mean trajectory of the bootstrap sample based on the virtual experimental data and the mean trajectory of the experiment.
In this way we obtain an empirical distribution of the $L^2$-norm distance due to natural fluctuation inside the experiment.
This is then compared with the $L^2$-norm distance of average trajectories between the experiment and the SFM and GCFM model.
The same procedure is also carries through for the Hilbert-Schmidt distance between the estimated correlation functions, see Figure~\ref{fig:HS_norm_EXP_GCFM_SFM_x}. 

Again, $p$-values are calculated as in the previous subsection, however this time we have to use one-sided regions of confidence for statistical testing.
The $p$-values are displayed in Table~\ref{tab:P_value_L_HS_norm}.

\begin{table}[h]
  \centering
  {
    \begin{tabular}{|r|r|r|}
      \hline
      \multicolumn{3}{|c|}{\textbf{p-values for $L^2$ norm}} \\
      \hline
   & EXP/GCFM & EXP/SFM \\
      \hline
      $x$-position   & 0     &  0   \\
      $y$-position   & 0.512 & 0.576\\
      \hline
    \end{tabular}
    ~~
    \begin{tabular}{|r|r|r|}
      \hline
      \multicolumn{3}{|c|}{\textbf{p-values for HS norm}} \\
      \hline
   & EXP/GCFM & EXP/SFM \\
      \hline
      $x$-position   & 0.122  & 0.434 \\
      $y$-position   & 0      & 0     \\
      \hline
    \end{tabular}
  }
  \caption{The p-values for $L^2$ norm and Hilbert-Schmidt norm .}
  \label{tab:P_value_L_HS_norm}
\end{table}

All statistical tests of the $L^2$-norm distance between the average $x$-values and $v_x$-values of the experiments and the models are highly significant.
We find thus a clear indication that model and experiment are statistically distinguished.
Not unexpectedly, the situation is different for the average of the $y$-values over time.
Here, due to the axial symmetry of the experimental set up, no major differences of the average $y$-trajectory can be observed.
This also shows that the slight asymmetry in the average $y$-trajectory in the experimental data is not statistically significant.

With regard to the fluctuation structure, the $x$-position fluctuations encoded by the empirical correlation function can not be easily distinguished between the SFM model and the experiment.
The difference between the fluctuation of the $x$-trajectories of the GCFM and the experiment also shows a very marginal $p$-value of $\approx 12\%$, which is not significant when compared to the usual $5\%$-level of significance.
All other fluctuation structures significantly differ between the experiment and the models.   

\section{\label{sec:summary}Summary}
The functional PCA has been applied as a diagnostic tool to assess model quality for agent-based simulations of pedestrian flows with respect to average behavior and beyond.
Here we applied it to experimentally measured pedestrian trajectories passing through a bottleneck and agent trajectories simulated by two different force-based models. Both models are a-priori calibrated to satisfy qualitative and quantitative criteria. 

Already in the analysis of mean flow behavior, the PCA reveals considerable and statistically significant deviation of both models from the experiment.
In the $x$-direction, the SFM and the GCFM predict slower progress and lower velocities leading to a longer dwell time before the bottle neck, as experimentally observed, although total through flow is being produced correctly.
This effect is more pronounced for the GCFM average behavior so that the SFM model reproduces the average behavior of the experiment -- for the given set of parameters -- in a relatively better way.

Coming to the statistical variations simulated in the SFM for $x$-position, we find a quite reasonable match in the qualitative behavior (the PCA mode shapes) of the $x$- and $y$-position over time.
A certain deviation in the quantitative variation strength is observed as well.
Again the SFM predictions are a little closer, when measured in terms of the deviation in the total variation.
Also, the concentration of variability to dominating modes is under estimated by the SFM and over estimated by the GCFM for the $x$-variations, while in the velocity variations both models show a higher degree of concentration as compared with the experiment.

Summarizing the PCA gives none of the models a clear ``pass'', as the SFM and the GCFM both significantly differ from the experiment, although both models were validated quantitatively with respect to the experimental flow through the bottleneck.
The SFM however performs relatively better than the GCFM, which is mostly due to a gradually better prediction of the average $x$-positions and $x$-velocities.

Caution is needed when applying these (and presumably also other) models to evacuation studies, as the evidently do not capture all features of real life pedestrian flows, as has been shown by our functional data analysis.
The overall picture of all the qualitative metrics derived from the PCA however slightly favours the SFM as the more accurate model over the GCFM, given our set of model parameters.

In this case study, applying functional PCA for the first time to pedestrian flows, we have thus shown that it is in fact a useful tool to benchmark and statistically test agent based pedestrian flow models.
Given the amount of deviation between experiment and model, it is certainly of interest to use this methodology for the future refinement of pedestrian flow models and for the critical assessment of models used by practitioners.

\section*{Acknowledgments}
We would like to thank Stefan Rosbach for his preparatory work on the functional PCA analysis for pedestrian data.

\bibliographystyle{00}

\begin{thebibliography}{10}
  \providecommand{\urlprefix}{}
  \expandafter\ifx\csname urlstyle\endcsname\relax
  \providecommand{\doi}[1]{doi:\discretionary{}{}{}#1}\else
  \providecommand{\doi}{doi:\discretionary{}{}{}\begingroup
    \urlstyle{rm}\Url}\fi
  
  
  
  
\bibitem{Bauer} Bauer, H., Probability Theory, de Gruyter 1995.
  
  
\bibitem{Chraibi2010a}
  Chraibi, M., Seyfried, A., and Schadschneider, A., {The generalized centrifugal
    force model for pedestrian dynamics}, \emph{Phys. Rev. E} \textbf{82} (2010)
  046111.
  
\bibitem{Chraibi2015}
  Traffic and Granular Flow '13, Eds. M. Chraibi, M. Boltes, A. Schadschneider and A. Seyfried, Springer (2015)
  
\bibitem{DiaconisEfron1983} Diaconis, P. and Efron, B., Computer-intensive methods in statistics. Scientific American,
  {\bf 248} No. 5 (1983) 116--130.
  
\bibitem{FisherCaffoSchwartzZipunnikov2014} 
  Fisher, A., Caffo, B., Schwartz, B. and Zipunnikov, V.,
  Fast, Exact Bootstrap Principal Component Analysis for
  $p > 1$ million, arXiv:1405.0922v3.
  
\bibitem{Helbing2004}
  Helbing, D., {Collective phenomena and states in traffic and self-driven many-particle systems}, \emph{Comp. Mater. Sci.} \textbf{30} (2004) 180--187.
  
\bibitem{Helbing1995}
  Helbing, D. and Moln\'{a}r, P., {Social force model for pedestrian dynamics},
  \emph{Phys. Rev. E} \textbf{51} (1995) 4282--4286.
  
\bibitem{Hoogendoorn2006}
  Hoogendoorn, S.~P. and Daamen, W., {Microscopic Parameter Identification of
    Pedestrian Models and its Implications to Pedestrian Flow Modeling}
  (Transporttation Research Board Annual Meeting, 2006).
  
\bibitem{Jian2010}
  Jian, M., Weiguo, S., Jun, Z., Siuming, L., and Guahgxuan, L.,
  {k-Nearest-Neighbor interaction induced self-organized pedestrian counter
    flow}, \emph{Physica A: Statistical Mechanics and its Applications}
  \textbf{389} (2010) 2101--2117.
  
\bibitem{Johansson2007}
  Johansson, A., Helbing, D., and Shukla, P.~K.
  {Specification of the social force pedestrian model by evolutionary adjustment to video tracking data},
  \emph{Advances in Complex Systems} \textbf{10} (2007) {271--288}.
  
\bibitem{Karamouzas2014}
  Karamouzas, I., Skinner, B., Guy, S.~J,
  {A universal power law governing pedestrian interactions},
  \emph{Phys. Rev. Lett.} \textbf{113} (2014) 238701
  
\bibitem{Lakoba2005}
  Lakoba, T.~I., Kaup, D.~J., and Finkelstein, N.~M., {Modifications of the
    Helbing-Moln\'{a}r-Farkas-Vicsek social force model for pedestrian
    evolution}, \emph{Simulation} \textbf{81} (2005) 339--352.
\bibitem{LiaoW2014b}
  Liao, W., Chraibi, M., Seyfried, A.,  Zhang, J., Zheng, X. and Zhao, Y.
  {Validation of FDS+Evac for pedestrian simulations in wide bottlenecks}, \emph{Intelligent Transportation Systems (ITSC), 2014 IEEE 17th International Conference}, (2014)  554--559.
  
\bibitem{Molnar1995}
  Moln\'ar, P., {Modellierung und Simulation der Dynamik von  Fu{\ss}g\"angerstr\"omen},
  Ph.D. Thesis. University Stuttgart 1996.
  
\bibitem{Nishinari2006}
  Nishinari, K., Sugawara, K., and Kaza, {Modelling of self-driven particles:
    Foraging ants and pedestrians}, \emph{Physica A} \textbf{372} (2006)
  132--141.
  
\bibitem{Parisi2007}
  Parisi, D.~R. and Dorso, C.~O., {Morphological and dynamical aspects of the
    room evacuation process}, \emph{Physica A} \textbf{385} (2007) 343--355.

  \bibitem{ped14}
    Daamen, W., Duives, C. and Hoogendoorn, S. P. (eds.)
    \emph{The Conference on Pedestrian and Evacuation Dynamics 2014 (PED 2014)}
    
\bibitem{Pipes1953}
  Pipes, L.~A., An operational analysis of traffic dynamics, \emph{J. Appl.
    Phys.} \textbf{24} (1953) 274 -- 281.
  
\bibitem{RHG} Ramsay, J.,  Hooker, G.  and Graves, S.\, , Functional Data Analysis with R and Matlab, UseR! Series,
  Springer Dortrecht Heidelberg London New York, 2009.
  
\bibitem{RS} Ramsay, J. and Silverman, B., Functional Data Analysis, Springer Verlag New York, 2005.
  
\bibitem{Seyfried2008}
  Seyfried, A. and Schadschneider, A., {Fundamental Diagram and Validation of
    Crowd Models}, in \emph{Cellular Automata}, eds. Umeo, H., Morishita, S.,
  Nishinari, K., Komatsuzaki, T., and Bandini, S., \emph{Lecture Notes in
    Computer Science}, Vol. 5191/2008 (Springer, Berlin Heidelberg, 2008), pp.
  563--566, \doi{10.1007/978-3-540-79992-4}.
  
  
\bibitem{Seyfried2010}
  Seyfried, A.; Boltes, M.; Kähler, J.; Klingsch, W.; Portz, A.; Rupprecht, T.; Schadschneider, A.; Steffen, B. and Winkens,A.
  Klingsch, W. W. F.; Rogsch, C.; Schadschneider, A. and Schreckenberg, M. (Eds.){Enhanced empirical data for the fundamental diagram and the flow through bottlenecks}
  \emph{Pedestrian and Evacuation Dynamics} Springer Berlin Heidelberg (2010) 145-156
\bibitem{Seyfried2010a}
  Seyfried, A. and Schadschneider, A., {Empirical results for pedestrian dynamics
    at bottlenecks}, in \emph{Parallel Processing and Applied Mathematics}, eds.
  Wyrzykowski, R., Dongarra, J., Karczewski, K., and Wasniewski, J.,
  \emph{Lecture Notes in Computer Science}, Vol. 6068 (Springer, Berlin
  Heidelberg, 2010), pp. 575--584,
  \urlprefix\url{http://dx.doi.org/10.1007/978-3-642-14403-5_62}.
  
\bibitem{Seyfried2006}
  Seyfried, A., Steffen, B., and Lippert, T., {Basics of modelling the pedestrian
    flow}, \emph{Physica A} \textbf{368} (2006) 232--238.
  
\bibitem{Shiwakoti2011}
  Shiwakoti, N., Sarvi, M., Rose, G., and Burd, M., Animal dynamics based
  approach for modelling pedestrian crowd egress under panic conditions,
  \emph{Transportation and Traffic Theory} \textbf{17} (2011) 438--461.
  
\bibitem{Schneider2002}
  Schneider, V. and K\"onnecke, R. {Simulating evacuation processes with ASERI},
  \emph{Pedestrian and Evacuation Dynamics} Springer (2002) 303--314
  
  
\bibitem{Steffen2010a}
  Steffen, B. and Seyfried, A., {Methods for measuring pedestrian density, flow,
    speed and direction with minimal scatter}, \emph{Physica A} \textbf{389}
  (2010) 1902--1910.
  
\bibitem{TraffGo2005}
  TraffGo HT GmbH, {Handbuch PedGo 2, PedGo Editor 2}, \urlprefix\url{www.evacuation-simulation.com} (2005)
  
\bibitem{Yu2005}
  W. J. Yu, R. Chen, L. Y. Dong and S. Q. Dai, {Centrifugal force model for pedestrian
    dynamics}, \emph{Phys.\ Rev.\  E}, \textbf{72} (2005) p. 026112.
\bibitem{Zhang2011a}
  Zhang, J., Klingsch, W., Schadschneider, A., and Seyfried, A., Transitions in
  pedestrian fundamental diagrams of straight corridors and t-junctions,
  \emph{J. Stat. Mech.}  (2011).
  
\bibitem{ptv}
 \url{http://vision-traffic.ptvgroup.com/de/produkte/ptv-viswalk/}
 
\bibitem{legion}
 \url{http://www.legion.com/}
 
 
\end{thebibliography}

\end{document}